\documentclass{article}
\usepackage{amsmath}  
\usepackage{amssymb}  
\usepackage{amsfonts} 
\usepackage{amsthm}   
\usepackage{bm}       
\usepackage{mathtools} 
\usepackage{cancel}
\usepackage{comment}
\usepackage{xr-hyper}
\usepackage{hyperref}
\hypersetup{hidelinks}
\usepackage[dvipsnames]{xcolor}
\usepackage{enumitem}
\usepackage{booktabs}
\usepackage{multirow}
\usepackage{adjustbox}
\usepackage{tikz}
\usepackage{amsmath}
\usetikzlibrary{patterns, decorations.pathreplacing}
\usepackage{array}
\usepackage{makecell}
\usepackage{natbib}
\usepackage{bbm}
\usepackage{dsfont}
\usepackage{authblk} 
\usepackage{abstract}

\usepackage{algorithm}
\usepackage{algpseudocode}
\usepackage[normalem]{ulem} 
\usepackage[a4paper,margin=0.9in]{geometry}
\usepackage{setspace}
\newcommand{\E}{\mathbb{E}}
\newcommand{\V}{\mathbb{V}}

\newcommand{\red}{\textcolor{red}}

\newcommand{\Prob}{\mathbb{P}}
\newcommand{\R}{\mathbb{R}}

\newcommand{\had}{\circ}          
\DeclareMathOperator{\diag}{diag}

\newtheorem{theorem}{Theorem}
\newtheorem{lemma}{Lemma}
\newtheorem{corollary}{Corollary}
\newtheorem{proposition}{Proposition}
\newtheorem{remark}{Remark}

\newtheorem{assumption}{Assumption}

\newtheorem{assumptionB}{Assumption}

\theoremstyle{definition} 

\makeatletter
\def\@biblabel#1{}
\makeatother

\newcommand{\tl}{\xrightarrow{\hspace{0.1cm}\mathcal{L}\hspace{0.1cm}}}
\newcommand{\tp}{\xrightarrow{\hspace{0.1cm}\Prob\hspace{0.1cm}}}

\date{}

\title{Hypothesis testing for estimable parameters with missing data: available cases or complete cases?}

\begin{document}
\author[1]{M. Romero-Madro\~nal\thanks{Corresponding author. Email address: \texttt{mrmadronal@us.es}}}
\author[1]{M.D. Jim\'enez-Gamero}
\author[2]{B. Milošević}

\affil[1]{Department of Statistics and Operations Research, Universidad de Sevilla, Spain\\ Instituto de Matem\'aticas de la Universidad de Sevilla, Spain}
\affil[2]{Faculty of Mathematics, University of Belgrade, Serbia}






\maketitle

\begin{abstract}
Whether to base inferences on available-case (AC) or complete-case (CC) estimators is a long-standing practical dilemma when data are missing. This issue has been investigated for point estimation in specific problems, but hypothesis testing has received comparatively little attention. The tools needed to address this problem are developed in a fully nonparametric framework for hypotheses concerning parameters that can be estimated through functions of vectors of $U$-statistics, under a MCAR missingness mechanism. The joint asymptotic normality of the AC estimators is established, permitting availability probabilities to vary with the sample size. The limiting covariance matrix factorizes into the complete-data covariance matrix and a matrix determined solely by the availability pattern, with both components admitting consistent estimators. This fact enables the construction of a Wald-type test statistic. The CC approach arises as a special case. These results allow AC and CC based tests to be compared, although the comparison is problem specific, which limits its practical value. Rather than selecting one approach, a test based on the maximum of the two statistics is also considered, whose critical value can consistently approximated by two bootstrap procedures. In simulations, its power lies between those of the AC and CC approaches.

\end{abstract}


\begingroup
\small
\urlstyle{same}
\setlength{\parindent}{0pt}
\setlength{\parskip}{0.5em}
\vfill
\noindent\rule{\linewidth}{0.3pt}\par
\noindent\textbf{Data availability statement.}
The data that support the findings of this study are available in National Health and Nutrition Examination Survey (NHANES) at \url{https://wwwn.cdc.gov/nchs/nhanes/continuousnhanes/default.aspx?BeginYear=2017}. These data were derived from the following resources available in the public domain:
\begin{itemize}[leftmargin=1.2em,topsep=0pt,itemsep=0pt,parsep=0pt]
\item National Health and Nutrition Examination Survey (NHANES),\\
\url{https://wwwn.cdc.gov/nchs/nhanes/continuousnhanes/default.aspx?BeginYear=2017}
\end{itemize}
\noindent\textbf{Conflict of interest statement.}
The authors declare no conflict of interest.
\par
\endgroup
\clearpage

\section{Introduction}

Missing data problems are commonly encountered in data analysis (\cite{little2019}). The way in which incomplete observations are handled  has important consequences for the efficiency  and the validity of statistical inference.

In multivariate studies, missingness is frequently componentwise: an observation may be available for some variables but unavailable for others. Consequently, different parameters may be estimated from different subsets of the original sample. A common way of dealing with missing data is complete-case (CC), or listwise deletion, which consists of carrying out inference from the set of complete cases. Although simple to implement, this approach may discard a substantial amount of usable information, particularly when the number of variables is moderate or large. 
Other commonly used approaches include single and multiple imputation, which require additional model assumptions 
\citep{vanbuuren2018, little2019, ramosaj2020}.

An alternative is the available-case (AC), in which each parameter is estimated using all  observations for which the components required for its calculation are available. This approach can potentially avoid a substantial loss of information compared to CC analysis, but it introduces an important theoretical difficulty. Different estimators are then based on different, randomly overlapping subsets of the sample.

The choice between these two procedures is  a long-standing practical dilemma. Much of the literature focuses on the estimation of the parameter vector in the linear regression model. Some studies assume normality \citep{wilks1932,matthai1951,glasser1964,haitovsky1968}, whereas others do not \citep{VanPraag1985,Xu2022Demystifying}. \cite{Xu2022Demystifying} compares the AC and CC estimators of the regression coefficient vector and shows that, for certain parameter values, the CC estimator may outperform the AC estimator in terms of having a smaller asymptotic variance. However, to the best of our knowledge, this comparison has not been addressed in the context of hypothesis testing. 

As a first step towards such a comparison, we establish general asymptotic results for both procedures. Deriving them requires some care because, under both AC and CC, the effective sample size is a random variable, and thus standard multivariate asymptotic theory, which typically relies on a common deterministic sample size, cannot be applied. These issues become particularly relevant for statistics constructed from several $U$-statistics, since estimators and test statistics often admit such representation. While the asymptotic theory of $U$-statistics for complete data is well established (see e.g. \cite{Hoeffding1948,lee1990}), the existing literature on $U$-statistics with missing observations has mainly considered some univariate parameters 
(\cite{aleksic2023}), specific multivariate parameters (\cite{VanPraag1985}, \cite{aleksic2025bhep}, \cite{gaigall2026bhep}), or assumed that the data has some specified distribution (\cite{Krishnamoorthy1998,Yu2006} for inferences on the mean of multivariate normal populations; \cite{HAO200162,JAMSHIDIAN20074227}  for inferences on the covariance matrix of multivariate normal populations), which may usually be unrealistic in practical applications.

 We start by establishing the joint asymptotic normality of the AC estimators with component-specific normalizations under missing completely at random (MCAR). This study is conducted in a fully nonparametric framework, that is, no assumptions are made regarding the population distribution, except for the existence of moments of certain functions of the data. The relevant kernels need not have the same degree or involve the same variables, and the corresponding availability probabilities need not be fixed. The asymptotic distribution of the CC estimator is also derived, as it arises as a special case of the general framework assumed for the AC case. This lets us to compare the AC and CC estimators from an asymptotic perspective within a general setting, which encompasses the parameter considered in \cite{Xu2022Demystifying}.

In order to apply the obtained results to the testing problem, an estimator of the asymptotic variance is required. This variance can be expressed as the product of several matrices, one of which depends explicitly on the missingness mechanism. In the absence of missing data, the results of \cite{Arvesen1969}  show that the jackknife provides a consistent estimator of the variance. We show, however, that when applied to incomplete data, the jackknife fails to consistently estimate the factor associated with the missingness mechanism. Nevertheless, this factor can be estimated separately, which allows us to construct a consistent estimator of the asymptotic variance and, consequently, to carry out valid inference.

The preceding results let us derive AC and CC Wald-type test statistics for linear hypotheses on the parameters of a single population, as well as compare their powers. This framework also extends to testing the equality of a parameter across multiple populations. 
In each case, the optimal choice between AC and CC depends on unknown population quantities, making it difficult to determine which procedure should be preferred in practice. To avoid this selection problem, we also consider the maximum of these test statistics. Since its asymptotic distribution does not possess a closed-form expression, we propose two consistent bootstrap estimators of its null distribution.

The finite-sample performance of these tests is assessed through simulation in two specific instances. Specifically, we consider testing the equality of the variances of the components of a $d$-variate vector and the equality of the coefficients of variation of each component of a $d$-variate vector (with positive components) across several populations.

The remainder of the paper is organized as follows. In Section \ref{sec:setting} we introduce the statistical setting and formally define the AC and CC estimators. Section \ref{sec:factorization} derives the asymptotic distribution. Section \ref{sec:variance} is devoted to variance estimation. Section \ref{sec:tests} builds AC and CC based Wald-type test statistics for hypothesis on a single population and for comparing parameters along $k$ populations, both tests are consistent against fixed alternatives and their asymptotic power may differ against local alternatives. Section \ref{sec:comparison} compares their powers. As the optimal choice depends on unknown quantities, Section \ref{subsec:boot} studies a test whose test statistic is the maximum of the AC and CC Wald statistics, and proposes two consistent approximations to its null distribution: a bootstrap in the limit and a multiplier bootstrap. Section \ref{sec:applications} summarizes the results of an extensive simulation experiment, designed to assess the finite-sample performance of the studied tests, both for synthetic data and  real data sets. Finally, Section \ref{sec:conclusions} concludes and underlines future research. All proofs and other technical details are provided in the Supporting Information.

\section{Statistical setting}\label{sec:setting}

Let
$X=(X^{(1)},\ldots,X^{(p)})^{\top} \in \bigotimes_{\ell=1}^p \mathcal{E}_{\ell} \subseteq \R^p,$ be a random vector and let $\bm X$ be a random sample from $X$, that is,
\begin{equation}
\label{eq:sampling_scheme}
\bm X = \{X_1,\ldots,X_n\},
\qquad
X_1,\ldots,X_n \text{ are independent copies of } X.
\end{equation}

Let \(h=(h^{(1)},\ldots,h^{(q)})^\top\) be a vector of symmetric kernels with degrees
\(m=(m_1,\ldots,m_q)\), and let \(A_r\subseteq\{1,\ldots,p\}\), \(1\leq r\leq q\),
denote the set of components of $X$ involved in the definition of \(h^{(r)}\). Accordingly,
\[
h^{(r)}: \mathcal{E}_{A_r}^{m_r} \longrightarrow\mathbb R,
\qquad
\mathcal{E}_{A_r}=\bigotimes_{\ell\in A_r} \mathcal{E}_\ell .
\]
We write
\begin{equation}
\label{eq:subvector_kernel_components}
X_{j}^{A_r}=\left(X_{j}^{(\ell)}\right)_{\ell\in A_r},
\qquad 1\leq j\leq n,\; 1\leq r\leq q,
\end{equation}
the subvector of \(X_{j}\) containing exactly the components required to evaluate the
\(r\)-th kernel. The parameters of interest are
\[
\eta=(\eta^{(1)},\ldots,\eta^{(q)})^\top,
\qquad
\eta^{(r)}=\mathbb E\left\{h^{(r)}\left(X_{1}^{A_r},\ldots,X_{m_r}^{A_r}\right)\right\},
\quad 1\leq r\leq q,
\]
and the $U$-statistic estimator of \(\eta^{(r)}\) is
\begin{equation}
\label{eq:complete_u_statistic}
\widehat\eta^{(r)}
=\binom{n}{m_r}^{-1}
\sum_{1\leq j_1<\cdots<j_{m_r}\leq n}
h^{(r)}\left(X_{j_1}^{A_r},\ldots,X_{j_{m_r}}^{A_r}\right),
\quad 1\leq r\leq q,
\end{equation}
with \(\widehat\eta=(\widehat\eta^{(1)},\ldots,\widehat\eta^{(q)})^\top\).  

Sometimes, the data will be assumed to form a triangular array, that is,
\begin{equation}
\label{eq:sampling_scheme_triangular}
\bm X_n = \{X_{n,1},\ldots,X_{n,n}\},
\qquad
X_{n,1},\ldots,X_{n,n} \text{ are independent copies of } X_n,
\end{equation}
where $X_n=(X_n^{(1)},\ldots,X_n^{(p)})^{\top}$ is a random vector taking values in $\R^p$, whose distribution may vary with $n$. In this setting, we write $\eta_n$ and $\widehat{\eta}_n$ instead of $\eta$ and $\widehat{\eta}$, respectively. The same convention applies for other quantities of interest.

Along this paper, unless otherwise stated, it will be assumed that the data is a sequence of independent and identically distributed (i.i.d) data, that is, that \eqref{eq:sampling_scheme} holds. All limits
are understood as \(n\to\infty\), and $p$ and $q$ are fixed quantities.

\subsection{Complete-data asymptotics}\label{subsec:complete_asymptotics}

For $1\leq r\leq q$, define the first-order projection by
\[
h_1^{(r)}\left(x^{A_r}\right)
=\mathbb E\left\{
h^{(r)}\left(X_1^{A_r},\ldots,X_{m_r}^{A_r}\right)
\,\middle|\,X_1^{A_r}=x^{A_r}
\right\}.
\]
Let \(\Gamma=(\Gamma_{rs})_{1\leq r,s\leq q}\) be the \(q\times q\) matrix with
entries
\begin{equation}
    \Gamma_{rs}
=\operatorname{Cov}\left\{h_1^{(r)}\left(X^{A_r}\right), h_1^{(s)}\left(X^{A_s}\right)\right\},
\qquad 1\leq r,s\leq q .
\label{eq:gamma}
\end{equation}
\begin{assumption}\label{as:moments}
\(\mathbb E\big\{h^{(r)}(X_{1}^{A_r},\ldots,X_{m_r}^{A_r})^2\big\}<\infty\), 
\(1\leq r\leq q\).
\end{assumption}
\begin{assumption}\label{as:nondeg}
\(\Gamma_{rr}=\V\{h_1^{(r)}(X^{A_r})\}>0\), \(1\leq r\leq q\).
\end{assumption}
Under Assumptions~\ref{as:moments}--\ref{as:nondeg}, Theorem 7.1 of \cite{Hoeffding1948} shows that
\begin{equation}
\sqrt{n}\left(\widehat\eta-\eta\right)\tl \mathcal N_q(0,V),
\qquad
V=\left(m_r m_s\,\Gamma_{rs}\right)_{1\leq r,s\leq q}.
\label{eq:eta_asymptotic_distribution}
\end{equation}

\subsection{Missingness mechanism and the AC estimator}\label{subsec:mcar}

Let \(M_{j}^{(\ell)}\in\{0,1\}\) denote the missingness indicator for the component
\(X_{j}^{(\ell)}\), where \(M_{j}^{(\ell)}=1\) if \(X_{j}^{(\ell)}\) is observed
and \(M_{j}^{(\ell)}=0\) otherwise. Define the augmented observations
\begin{equation}
\label{eq:augmented_observations}
Z_{j}=
\left(
X_{j}^{(1)},\ldots,X_{j}^{(p)},
M_{j}^{(1)},\ldots,M_{j}^{(p)}
\right),
\qquad 1\leq j\leq n.
\end{equation}

\begin{assumption}\label{as:mcar}
 For each $n$, the augmented observations \(Z_{1},\ldots,Z_{n}\) are  i.i.d., and the missingness mechanism is MCAR, i.e.
\begin{equation}
\label{eq:mcar}
\left(M_{j}^{(1)},\ldots,M_{j}^{(p)}\right)\ \text{is independent of}\ X_{j},
\qquad 1\leq j\leq n .
\end{equation}
\end{assumption}
Notice that no independence among the components \(M_{j}^{(1)},\ldots,M_{j}^{(p)}\) is assumed.

Since \(h^{(r)}\) depends only on the components indexed by \(A_r\), it is enough to
retain observations for which those components are observed. Define the
availability indicator
\begin{equation}
R_{j}^{(r)}=\prod_{\ell\in A_r}M_{j}^{(\ell)},
\qquad 1\leq j\leq n,\; 1\leq r\leq q,
\label{eq:availability_indicator}
\end{equation}
so that \(R_{j}^{(r)}=1\) iff all components required by \(h^{(r)}\) are observed for
the \(j\)-th observation. We denote by
\[
N^{(r)}=\sum_{j=1}^{n}R_{j}^{(r)}
\]
the random number of observations available for the computation of the
\(U\)-statistic associated with \(h^{(r)}\).
The AC estimator of \(\eta^{(r)}\) is
\begin{equation}
\widehat\eta_{AC}^{(r)}
=\binom{N^{(r)}}{m_r}^{-1}
\sum_{1\leq j_1<\cdots<j_{m_r}\leq n}
h^{(r)}\left(X_{j_1}^{A_r},\ldots,X_{j_{m_r}}^{A_r}\right)
\prod_{a=1}^{m_r}R_{j_a}^{(r)},
\label{eq:available_case_estimator}
\end{equation}
whenever $N^{(r)}\geq m_r,$ with \(\widehat\eta_{AC}=(\widehat\eta_{AC}^{(1)},\ldots,\widehat\eta_{AC}^{(q)})^\top\).

The AC estimators are conditionally unbiased given the missingness
pattern under MCAR as shown in the next result. 

\begin{proposition}\label{prop:cond_unbiased}
Suppose that Assumption~\ref{as:mcar} holds. Then, on the event \(\{N^{(r)}\ge m_r\}\),
\[
\E\big(\widehat\eta_{AC}^{(r)}\,\big|\, R_1^{(r)},\ldots,R_n^{(r)}\big)=\eta^{(r)},
\qquad 1\le r\le q.
\]

\end{proposition}
As an immediate consequence of Proposition \ref{prop:cond_unbiased}, the AC estimator is unbiased.
\begin{corollary}
\label{corollary:unbiased}
    Suppose that the assumptions in Proposition \ref{prop:cond_unbiased} hold. Then on the event \(\{N^{(r)}\ge m_r\}\), $\E\big(\widehat\eta_{AC}^{(r)}\big)=\eta^{(r)},
$ $1\le r\le q$ . 
\end{corollary}
For \(1\leq r\leq q\), define the availability probabilities
\begin{equation}
\pi_n^{(r)}=\Prob\left(R_{j}^{(r)}=1\right).
    \label{eq:pis}
\end{equation}
The subindex \(n\) indicates that that these probabilities may vary with $n$.

\begin{assumption}\label{as:avail}
$n\pi_n^{(r)}\to\infty,
$ $1\leq r\leq q .$
\end{assumption}

Recall that Proposition \ref{prop:cond_unbiased} and Corollary \ref{corollary:unbiased} hold on the event $\{N^{(r)}\ge m_r\}$. Next result shows that such event occurs with probability tending to 1 under Assumption \ref{as:avail}.  

\begin{proposition} \label{prop:event}
Suppose that Assumptions \ref{as:mcar}--\ref{as:avail} hold. Let \(E_n^{(r)}=\{N^{(r)}\ge m_r\}\) and  \(E_n=\bigcap_{r=1}^{q}E_n^{(r)}\). Then, $\Prob(E_n) \to 1$. 
\end{proposition}

Next we study the case $n \pi_n^{(r)} \to c \in \R$, $c\geq 0$.

\begin{proposition}
\label{prop:bad_case}
(a) If $n\pi_n^{(r)} \to 0$, for some $r\in\{1,\ldots, q\}$,  then $\Prob(E_n) \to 0$.
\newline (b) If $n\pi_n^{(r)} \to c>0$,  then  $\Prob(E^{(r)}_n) \to p_c<1$, where $p_c=\Prob(Y\geq m_r)$, $Y \sim Pois(c)$.
\end{proposition}

Proposition~\ref{prop:bad_case} shows that, if $n\pi_n^{(r)}$ converges to a finite limit, the probability that $\widehat\eta_{AC}^{(r)}$ cannot be calculated does not vanish. The asymptotic inference developed below therefore assumes that each expected available sample size diverges, as required by Assumption~\ref{as:avail}.

For \(1\leq r,s\leq q\) define the  joint availability probabilities
\begin{equation}
\pi^{(r,s)}_n=\Prob\left(R_{j}^{(r)}=1,R_{j}^{(s)}=1\right).
\label{eq:pis_joint}
\end{equation}
 Note that
\(\pi_n^{(r,s)}=\E(R_j^{(r)}R_j^{(s)})=\Prob\big(\prod_{\ell\in A_r\cup A_s}M_j^{(\ell)}=1\big)\),
so the joint availability is governed by the union \(A_r\cup A_s\) of the two
index sets. 
It is convenient to introduce the   quantities
\begin{equation}
\varrho^{(r,s)}_n=\frac{\pi^{(r,s)}_n}{\sqrt{\pi^{(r)}_n\pi^{(s)}_n}},
\qquad 1\leq r,s\leq q.
\label{eq:varrho}
\end{equation} 
The Cauchy--Schwarz inequality gives
\(\pi_n^{(r,s)}=\E(R_j^{(r)}R_j^{(s)})\le\sqrt{\E(R_j^{(r)})\,\E(R_j^{(s)})}=\sqrt{\pi_n^{(r)}\pi^{(s)}_n}\),
whence
\begin{equation*}
0\le \varrho^{(r,s)}_n\le 1,
\qquad \varrho^{(r,r)}_n=1.
\end{equation*}
Thus \(\varrho_n=(\varrho^{(r,s)}_n)_{1\le r,s\le q} \) is a genuine correlation-type matrix of the availability
 pattern. 
It carries the entire effect of missingness on the
limiting covariance of $\widehat{\eta}_{AC}$. To obtain a single limiting covariance matrix, we assume that it stabilizes.

\begin{assumption}\label{as:coupling}
The limit
\(\varrho^{(r,s)}:=\lim_{n\to\infty}\varrho_{n}^{(r,s)}\)
exist for all \(1\le r,s\le q\). We write
\(\varrho:=\left(\varrho^{(r,s)}\right)_{1\le r,s\le q}\).
\end{assumption}

\begin{remark} \label{rem:CC1}
As observed in the Introduction, a natural alternative for $\widehat\eta_{AC}$ commonly used in practice, is the CC estimator, which discards every observation with at least one relevant variable missing. Let $A_{full}=A_1\cup\cdots\cup A_q$ and
\[
R_j^{full}=\prod_{\ell\in A_{full}}M_j^{(\ell)},\qquad N_{full}=\sum_{j=1}^{n}R_j^{full},
\]
so that $R_j^{full}=1$ if and only if all the variables involved in the problem are available for the $j$-th observation. The CC estimator $\widehat\eta_{CC}$ is the vector of classical $U$-statistics computed on the subsample $\{1\leq j\leq n:\ R_j^{full}=1\}$. Note that $\widehat\eta_{CC}$ is the particular case of $\widehat\eta_{AC}$ obtained by formally enlarging every index set $A_r$ to $A_{full}$. As a consequence, all the results obtained for the AC case remain valid for the CC case. Let $\pi^{full}_n=\Prob(R^{full}_j=1).$
\end{remark}

\section{Asymptotic distribution}\label{sec:factorization}

This section shows that the asymptotic behavior based on the AC coincides with that of the complete-data estimator in \eqref{eq:eta_asymptotic_distribution}, with 
$\sqrt{n}$ and $V$ replaced with 
adequate quantities  that account for  the missingness mechanism.

\begin{theorem}\label{thm:sparse}
Suppose that Assumptions~\ref{as:moments}-\ref{as:coupling} hold. Then, $\mathcal D_n\big(\widehat\eta_{AC}-\eta\big)
\ \tl\ \mathcal N_q\big(0, V\had \varrho \big),$ where $\mathcal D_n
=\diag\Big(\big(n\,\pi_n^{(1)}\big)^{1/2},\ldots,
\big(n\,\pi_n^{(q)}\big)^{1/2}\Big)$,  \(\had\) denotes the Hadamard (componentwise) product, $V$ is the complete-data covariance of
\eqref{eq:eta_asymptotic_distribution} and $\varrho$ is defined in Assumption \ref{as:coupling}.
\end{theorem}

Under missing data, the available sample sizes are the random variables $N^{(r)}$, $1\leq r \leq q$. Note that the elements of the normalizing matrix $\mathcal{D}_n$  are the square roots of the expected values of these random available sample sizes. In practice, the result in Theorem \ref{thm:sparse} is infeasible as the true values of $\pi_n^{(r)}$, $1\leq r \leq q$, are often unknown. 
 Nevertheless, the next result shows that if $\mathcal{D}_n$ is replaced with the observed available sample size, we get the same limit distribution.
\begin{corollary}
\label{cor:1}
    Suppose that assumptions in Theorem \ref{thm:sparse} hold. Then,
$\widehat{\mathcal D}_n\big(\widehat\eta_{AC}-\eta\big)
\ \tl\ \mathcal N_q\big(0, V\had \varrho \big),$ where $\widehat{\mathcal D}_n
=\diag\Big(\big(N^{(1)}\big)^{1/2},\ldots,
\big(N^{(q)}\big)^{1/2}\Big)$.
\end{corollary}

Note that in Theorem   \ref{thm:sparse}  and Corollary \ref{cor:1} each component of $\hat{\eta}$ may have a different normalizing constant, that is, the elements of ${\mathcal D}_n$ (or  $\widehat{\mathcal D}_n$) may converge to $\infty$ at very different rates. Whenever they are comparable, in the sense stated in 
the next assumption, a common normalizing constant can be used.

\begin{assumption}\label{as:comparable}
$\pi_n^{(r)}/\sum_{s=1}^q \pi_n^{(s)}\to\kappa_r \in (0,1],
$ $1\leq r\leq q .$
\end{assumption}

\begin{corollary}
\label{cor:equal}
Suppose that  Assumptions~\ref{as:moments}-\ref{as:comparable} hold. Let $N=n\sum_{s=1}^q \pi_n^{(s)}$ and $\widehat{N}=\sum_{s=1}^qN^{(s)}$. Then,
\begin{enumerate}
\item $\sqrt{N}\left(\widehat\eta_{AC}-\eta\right)\tl
\mathcal N_q\!\left(0,V\had M\right)$, where $M= K\varrho  K$, $K=\diag (1/\sqrt{\kappa_1}, \ldots, 1/\sqrt{\kappa_q})$,
\item $\sqrt{\widehat{N}}\left(\widehat\eta_{AC}-\eta\right)\tl
\mathcal N_q\!\left(0,V\had M\right)$.
\end{enumerate}
\end{corollary}

Notice that if  $\pi_n^{(r)}=\pi^{(r)}\in(0,1]$, $1 \leq r \leq q$, then Assumption \ref{as:comparable} is satisfied. In such a case the normalizing constant can be chosen equal to $\sqrt{n}$.  Note that Theorem 2 of  \cite{aleksic2023}  is a special case for $q=1$. 


Some statistics of interest are smooth functions of a vector of $U$-statistics. Examples are the correlation coefficient,  the Gini index, and Kendall's tau. The next corollary shows that the previous results also apply to this class of statistics. 

\begin{corollary}
\label{cor:functions} 
    Suppose that  assumptions in Corollary~\ref{cor:equal} hold, that
     $f:\mathbb{R}^q \to \mathbb{R}^d$ is continuous at $\eta$ and  has nonzero differential at $\eta$, $\nabla f(\eta)$. Then,
     \begin{enumerate}
\item $\sqrt{N}
\left\{f(\widehat\eta_{AC})-f(\eta)\right\}\tl
\mathcal N_d\!\Big(0,  \nabla f(\eta) (V \had M ) \nabla f(\eta)^\top \Big),$
\item  $\sqrt{\widehat{N}}
\left\{f(\widehat\eta_{AC})-f(\eta)\right\}\tl
\mathcal N_d\!\Big(0,  \nabla f(\eta) ( V \had M)\nabla f(\eta)^\top \Big).$
\end{enumerate}
\end{corollary}

\begin{remark}\label{rem:cc}
As observed in Remark \ref{rem:CC1}, the above
results  also cover the CC estimator. They apply with with every index set enlarged to $A_{full}$. The assumptions then involve the single availability sequence $R_j^{full}$ and are otherwise unchanged.  All the availability sets coincide, so $N^{(r)}=N_{full}$ for every $r$, $\mathcal D_n=(n\pi^{full}_n)^{1/2}I_q$ and $\widehat{\mathcal D}_n=N_{full}^{1/2}I_q$, where $I_q$ denotes the $q \times q$ identity matrix, and $\varrho$ is the matrix of ones. Then from Theorem \ref{thm:sparse} we have that
\[
\sqrt{n\pi_n^{full}}\big(\widehat\eta_{CC}-\eta\big)\ \tl\ \mathcal N_q(0,V),
\]
and Corollary~\ref{cor:1} then reduces to
\[
\sqrt{N_{full}}\big(\widehat\eta_{CC}-\eta\big)\ \tl\ \mathcal N_q(0,V),
\]
that is, the complete-data limit \eqref{eq:eta_asymptotic_distribution} with the sample size replaced by the number of complete  cases. 
\end{remark}

The results above remain valid, under certain additional assumptions, when the data form a triangular array, that is, when \eqref{eq:sampling_scheme_triangular} holds. We now consider this setting. First, instead of the augmented observations $Z_j$ defined in \eqref{eq:augmented_observations}, we consider the following.
\begin{equation*}
Z_{n,j}=
\left(
X_{n,j}^{(1)},\ldots,X_{n,j}^{(p)},
M_{j}^{(1)},\ldots,M_{j}^{(p)}
\right),
\qquad 1\leq j\leq n.
\end{equation*}
Likewise, in this case, we replace Assumption \ref{as:mcar} by the following.
\begin{assumptionB}
\label{as:B1}
 For each $n$, the augmented observations \(Z_{n,1},\ldots,Z_{n,n}\) are  i.i.d., and the missingness mechanism is MCAR, i.e.
\begin{equation*}
\left(M_{j}^{(1)},\ldots,M_{j}^{(p)}\right)\ \text{is independent of}\ X_{n,j},
\qquad 1\leq j\leq n .
\end{equation*}
\end{assumptionB}

\begin{assumptionB}
\label{as:B2} 
Assume that \eqref{eq:sampling_scheme_triangular} holds and that there exist $\eta \in \mathbb{R}^q$ and $\Gamma \in \mathbb{R}^{q \times q}$ such that  
$\eta_n \to \eta$ and $\Gamma_n \to \Gamma$.
\end{assumptionB}

\begin{assumptionB}
\label{as:B3} 
    Assume that \eqref{eq:sampling_scheme_triangular} holds and that there exists $\delta>0$ such that \[\sup_n \E \big|h^{(r)}\big(X_{n,1}^{A_r},\dots,X_{n, m_r}^{A_r}\big)\big|^{2+\delta}<\infty, \quad  1\le r\le q.\]
\end{assumptionB}

\begin{theorem}
\label{prop:triangular}
Suppose that Assumptions \ref{as:B1}-- \ref{as:B3} hold, that $\Gamma$ in \ref{as:B2} satisfy Assumption \ref{as:nondeg},  and that Assumptions \ref{as:avail}, \ref{as:coupling} hold. Let $V=(m_rm_s\Gamma_{rs})_{1\le r,s\le q}$, and let $\mathcal D_n$, $\widehat{\mathcal D}_n$, $N$, $\widehat N$ and $M$ be as in Theorem \ref{thm:sparse} and Corollaries \ref{cor:1} and \ref{cor:equal}. Then,
\begin{enumerate}
\item \textup{(Theorem \ref{thm:sparse} and Corollary \ref{cor:1})} $\mathcal D_n\big(\widehat\eta_{n,AC}-\eta_n\big)\tl\mathcal N_q(0,V\had\varrho)$ and $\widehat{\mathcal D}_n\big(\widehat\eta_{n,AC}-\eta_n\big)\tl\mathcal N_q(0,V\had\varrho)$.
\item \textup{(Corollary \ref{cor:equal})} If, in addition, Assumption \ref{as:comparable} holds, then $\sqrt N\big(\widehat\eta_{n,AC}-\eta_n\big)\tl\mathcal N_q(0,V\had M)$ and $\sqrt{\widehat N}\big(\widehat\eta_{n,AC}-\eta_n\big)\tl\mathcal N_q(0,V\had M)$.
\item \textup{(Corollary \ref{cor:functions})} If, in addition to the assumptions in part 2, $f:\R^q\to\R^d$ has continuous first-order partial derivatives in a neighborhood of $\eta$ and nonzero differential at $\eta$, then
$\sqrt N\big\{f(\widehat\eta_{n,AC})-f(\eta_n)\big\}\tl\mathcal N_d\big(0,\nabla f(\eta)(V\had M)\nabla f(\eta)^\top\big)$ and $\sqrt{\widehat N}\big\{f(\widehat\eta_{n,AC})-f(\eta_n)\big\}\tl\mathcal N_d\big(0,\nabla f(\eta)(V\had M)\nabla f(\eta)^\top\big)$.
\end{enumerate}
\end{theorem}

\section{Variance estimation}\label{sec:variance}

To make inferences, we need to estimate the asymptotic covariance matrices appearing in the results of the previous section. Note that, in all cases, each covariance matrix can be expressed as the product of several factors. Therefore, if a consistent estimator is available for each factor, their product is a consistent estimator of the target matrix. Consistent estimators of 
 $\nabla f(\eta)$, $M$ and $\varrho$ are readily obtained by plug-in through  $\nabla f(\widehat{\eta}_{AC})$, $\widehat{M}=\widehat{K}\widehat{\varrho}\widehat{K}$ and $\widehat{\varrho}$,
 $$\widehat{K}=\diag (1/\sqrt{\widehat{\kappa}_1}, \ldots, 1/\sqrt{\widehat{\kappa}_q}), \quad \widehat{\kappa}_r=\frac{N^{(r)}}{\widehat N}, \quad \widehat{\varrho}=(\widehat{\varrho}^{(r,s)}), \quad \widehat{\varrho}^{(r,s)}=\frac{N^{(r,s)}}{\sqrt{N^{(r)}N^{(s)}}},$$
 where $N^{(r,s)}=\sum_{j=1}^{n}R_{j}^{(r)} R_{j}^{(s)}$ is the random number of observations available for the computation of the
\(U\)-statistics associated with \(h^{(r)}\) and \(h^{(s)}\).
 The main difficulty lies in estimating $V$.

In some particular cases, explicit expressions for $V$ are available. The covariance matrix can then be estimated by plugging in consistent estimators of the parameters involved (see, for example, \cite{VanPraag1985} for the estimation of the coefficient vector in the linear regression model). However, in many practical cases, $V$ does not admit an easily computable expression.

For the case of complete data,  
the results of \cite{Arvesen1969} (see also \cite{articulo2}) show that the jackknife provides a consistent estimator of the variance (of the asymptotic distribution) of $\sqrt{n} \{f(\widehat{\eta})-f(\eta)\}$. Here we investigate the asymptotic behavior of such estimator with MCAR missing data. 

The presence of missing data introduces additional complications in the construction of the jackknife variance estimator. In order to appreciate them, we first recall the definition of such estimator in the complete-data setting. 
Let \( \hat{\eta}_{-i}\) denote the estimator $\hat{\eta}$ computed without the \(i\)-th observation. Then,
\[\widehat{var}_{jack}(\hat{\eta})=(n-1)\sum_{i=1}^n(\hat{\eta}_{-i}-\hat{\eta}_{(.)})(\hat{\eta}_{-i}-\hat{\eta}_{(.)})^\top, \quad \hat{\eta}_{(.)}= \frac{1}{n}\sum_{i=1}^n\hat{\eta}_{-i},\]
is a consistent estimator of the asymptotic variance of $\sqrt{n}(\hat{\eta}-\eta)$, that is, $\widehat{var}_{jack}(\hat{\eta})$ converges in probability to $V$.

When the data are incomplete, an observation may contribute to the estimation of one component of $\hat{\eta}_{AC}$, say $\hat{\eta}^{(r)}_{AC}$, but not to another, say $\hat{\eta}^{(s)}_{AC}$. Consequently, two approaches can be considered for defining $\widehat{var}_{jack}(\hat{\eta}_{AC})$: (a) restricting the analysis to CC, that is, observations with all variables available; or (b) defining the estimator componentwise. Under the latter approach,  $\widehat{var}_{jack}(\widehat{\eta}_{AC})_{rs}$ is computed using the observations available for both $\widehat{\eta}^{(r)}$ and $\hat{\eta}^{(s)}$. Specifically, let $D_r , D_s\subseteq \{1,\ldots,n\}$ denote the sets of indices corresponding to the observations used to estimate $\widehat{\eta}^{(r)}_{AC}$ and $\widehat{\eta}^{(s)}_{AC}$, respectively. Then, $\widehat{var}_{jack}(\hat{\eta}_{AC})_{rs}$ is calculated from the observations indexed by the intersection $D_{rs}=D_r\cap D_s$. Let $n_{rs}=|D_{rs}|$, $ 1 \leq r,s \leq q$.

As observed in Remark \ref{rem:CC1}, CC can be seen as a special case of AC. Because of this reason, here we consider the second approach, based of the available data for each pair of components. 

Instead of  $\eta$ we  first consider  the more general case in Corollary \ref{cor:functions} where the objective is to estimate $f(\eta)$, $f:\mathbb{R}^q \to \mathbb{R}^d$, with components $f=(f^{(1)}, \ldots, f^{(d)})^\top$. Let $D^f_r$ denote the set of indices corresponding to the observations used to estimate $f^{(r)}(\eta)$, that is, those used for calculating  $f^{(r)}( \hat\eta_{AC})$. Let $D^f_{rs}=D^f_r\cap D^f_s$, and $n^f_{rs}=|D^f_{rs}|$, the number of elements in the set $D^f_{rs}$, $1\leq r,s \leq d$.
For $i\in D$, $D\subseteq\big\{1,\ldots,n\big\}$, let $\widehat\eta^{(r)}_{AC,-i}\big(D\big)$ denote the estimator $\widehat\eta^{(r)}_{AC}$ computed from the observations indexed by $D\setminus\big\{i\big\}$, and let $\widehat\eta_{AC,-i}\big(D\big)$ denote the corresponding vector.   Let   $\nabla f(\eta) V \nabla f(\eta)^\top =(\sigma_{rs})$.

\begin{proposition} \label{prop:jack}
Let $1\leq r,s\leq d$. Suppose that  Assumptions~\ref{as:moments}-\ref{as:coupling} hold, that $\E(n^f_{rs})\to \infty$, that $f$ has nonzero differential at $\eta$  and  has continuous first-order partial derivatives in a neighborhood of $\eta$.
 Let
\begin{align*}
    \widehat{\sigma}_{rs}= &\big(n^f_{rs}-1\big)
\sum_{i\in D^f_{rs}}
\left[f_r\big\{ \widehat\eta_{AC,-i}(D^f_{rs})\big\}-f_r\big( \widehat\eta_{AC}\big)_{(.)}\right]
\left[f_s\big\{ \widehat\eta_{AC,-i}(D^f_{rs})\big\}-f_s\big( \widehat\eta_{AC}\big)_{(.)}\right], \\ & f_v\big( \widehat\eta_{AC}\big)_{(.)}  = \frac{1}{n^f_{rs}}\sum_{i\in D^f_{rs}} f_v\big\{ \widehat\eta_{AC,-i}(D^f_{rs})\big\}, \quad v=r,s.
\end{align*} 
Then, $\widehat{\sigma}_{rs}\tp \sigma_{rs}$.
\end{proposition}

As a consequence of Corollary \ref{cor:functions} and Proposition \ref{prop:jack}, it follows that the  jackknife variance estimator of the asymptotic variance of  $\sqrt{N}
\left\{f(\widehat\eta_{AC})-f(\eta)\right\}$ (or $\sqrt{\hat N}
\left\{f(\widehat\eta_{AC})-f(\eta)\right\}$) is not a consistent estimator, as it does not incorporates the matrix $M$, that accounts for the missingness mechanism. Nevertheless, as a consequence of the previous result, 
the jackknife can be used to estimate the matrix $V=(V_{rs})$.

\begin{assumption}
    $n \pi^{(r,s)} \to \infty$, $1 \leq r,s \leq q$.
    \label{as:jack}
\end{assumption}
\begin{corollary}
\label{cor:jack}
Let $1\leq r,s\leq q$. Suppose that  Assumptions~\ref{as:moments}-\ref{as:coupling}, \ref{as:jack} hold. 
Let
\begin{align*}
    \widehat{V}_{rs}&= (n_{rs}-1)
\sum_{i\in D_{rs}}
\left\{ \widehat\eta^{(r)}_{AC,-i}(D_{rs})-\widehat\eta^{(r)}_{AC(.)}\right\}
\left\{ \widehat\eta^{(s)}_{AC,-i} (D_{rs})-\widehat\eta^{(s)}_{AC(.)}\right\}, \\
    \widehat\eta^{(v)}_{AC(.)} &= \frac{1}{n_{rs}}\sum_{i\in D_{rs}} \widehat\eta^{(v)}_{AC,-i}(D_{rs}), \quad  v=r,s.
\end{align*} 
Then, $\widehat{V}_{rs}\tp V_{rs}$.
\end{corollary}

\begin{remark}
    To perform inference, a consistent estimator of $V\had\varrho$ is required. Under the assumptions of Corollary \ref{cor:jack}, $\widehat V_{rs}$ is a consistent estimator of $V_{rs}$. If, otherwise, $n\pi^{(r,s)}_n$ remains bounded, it can be easily checked that $\varrho^{(r,s)}=0$, whereas $\widehat V_{rs}$ stays bounded in probability. In such a case, $\widehat V_{rs}\,\widehat{\varrho}^{(r,s)}\tp 0$, as it should.
\end{remark}

Notice that, although consistent for $V$, $\widehat V$ may not be positive semidefinite in finite samples, as their components are calculated from different data sets (because $D_{rs}$  may differ from $D_{tw}$ for $\{r,s\} \neq \{t,w\}$). In simulations we proceed as follows: 
whenever $\lambda_{\min}(\widehat V)<0$, we replace $\widehat V$ by its projection onto the positive semidefinite cone, that is, if $ \widehat{V}=O D O^{\top}$, $D=\diag(\widehat{\lambda}_1, \ldots, \widehat{\lambda}_q)$, where $\widehat{\lambda}_1 \geq \cdots \geq \widehat{\lambda}_t \geq 0 > \widehat{\lambda}_{t+1}\geq \cdots \geq \widehat{\lambda}_{q}$ are the eigenvalues of $\widehat{V}$ and $O$ is an orthogonal matrix, the cited projection is $\widehat V_{+}=O \diag(\widehat{\lambda}_1, \ldots, \widehat{\lambda}_t, 0, \ldots, 0) O^{\top}$. 

Proposition \ref{prop:jack}, and therefore Corollary \ref{cor:jack}, remain valid when the data form a triangular array, that is, when \eqref{eq:sampling_scheme_triangular} holds. We now consider this setting.

\begin{proposition}\label{prop:jack_triangular}
Let $1\leq r,s\leq d$. Suppose that Assumptions \ref{as:B1}--\ref{as:B3} hold, that $\Gamma$ in Assumption \ref{as:B2} satisfies Assumption \ref{as:nondeg}, that Assumptions \ref{as:avail}, \ref{as:coupling} hold, that $\E(n^f_{rs})\to\infty$, and that $f$ has nonzero differential at $\eta$ and continuous first-order partial derivatives in a neighborhood of $\eta$, where $\eta$ is the limit in Assumption \ref{as:B2}. Let $\widehat\sigma_{rs}$ be as in Proposition \ref{prop:jack}, computed from $X_{n,1},\ldots,X_{n,n}$. Then, $\widehat\sigma_{rs}\tp\sigma_{rs}$. In particular, if Assumption \ref{as:jack} holds, then $\widehat V_{rs}\tp V_{rs}$, with $\widehat V_{rs}$ as in Corollary \ref{cor:jack}.
\end{proposition}

\section{One-sample and \texorpdfstring{$\boldsymbol{k}$}{k}-sample Wald-type test statistics}\label{sec:tests}

The results of Sections~\ref{sec:factorization} and~\ref{sec:variance} provide the key ingredients for inference: the joint asymptotic normality of the AC estimator and a consistent estimator of its asymptotic covariance. This section develops hypothesis tests for general linear constraints of the form $H_0\colon Cf(\eta)=0$, for some fixed matrix $C$. We first treat the one-sample problem and then extend the framework to the comparison of $k$ independent populations. Specifically, we will deal with Wald-type tests.

\subsection{One sample tests}\label{sec:onesample}

For a linear null hypothesis $H_0: Cf(\eta)=0$ vs $H_1: Cf(\eta) \neq 0$, $C$ being a $c\times d$ matrix of full row rank, the  statistic of the Wald test is
\begin{equation}\label{eq:app:wald}
T_n=\Big\{Cf\big(\widehat\eta_{AC}\big)\Big\}^\top
\Big(C\,\widehat\Omega_f C^\top\Big)^{-}
\Big\{Cf\big(\widehat\eta_{AC}\big)\Big\},
\end{equation}
where $(\cdot)^{-}$ denotes the Moore--Penrose pseudoinverse (see e.g. \cite[p.~24]{graybill1976linear}), \[\widehat\Omega_f=\nabla f(\widehat\eta_{AC})\widehat\Omega\nabla f(\widehat\eta_{AC})^{\top},\] and $\widehat{\Omega}=(\widehat{\Omega}_{rs})$, with 
\begin{equation}
    \widehat\Omega_{rs}=(\widehat V_{+})_{rs}\,\frac{N^{(r,s)}}{N^{(r)}N^{(s)}},
    \label{eq:Omega}
\end{equation}
where $\widehat V_{+}$ is the projected estimator defined after Corollary \ref{cor:jack}. Notice that $\widehat\Omega$ is  positive semidefinite as it is obtained as the Hadamard product of two  positive semidefinite matrices.

+For complete data, the Wald statistic has a standard chi-square limit under the usual regularity conditions. The incomplete-data setting considered here cannot be viewed as a special case of this general result, since the sample sizes are themselves random variables.  
The next result shows that, under some regularity conditions, $T_n$ has the same limit in law as in the complete-data setting. Let $\|\cdot \|$ denote the Euclidean norm.

 \begin{proposition}\label{prop:waldf}
Suppose that $f:\mathbb{R}^q \to \mathbb{R}^d$ is continuous at $\eta$, has nonzero differential at $\eta$, $\nabla f(\eta)$, and each component of $f$ has continuous first-order partial derivatives in a neighborhood of $\eta$. Let $\Sigma_f=C\,\nabla f(\eta)\big(V\had M\big)\nabla f(\eta)^\top C^\top$, $C$ being a $c\times d$ matrix of full row rank, and suppose that $\Sigma_f$ is non-singular and that either Assumption \ref{as:jack} holds or $n\pi^{(r,s)}_n$ remains bounded, $1\leq r,s\leq q$. Let $T_n$ be defined as in \eqref{eq:app:wald}.
\begin{enumerate}
\item If Assumptions \ref{as:moments}--\ref{as:comparable} hold and $Cf(\eta)=0$, then $T_n\tl\chi^2_c$.
\item If Assumptions \ref{as:moments}--\ref{as:comparable} hold and $Cf(\eta)\neq0$, then $T_n\tp\infty$.
\item If Assumptions \ref{as:B1}--\ref{as:B3} hold, $\Gamma$ in Assumption \ref{as:B2} satisfies Assumption \ref{as:nondeg}, Assumptions \ref{as:avail}--\ref{as:comparable} hold, and $N^{1/2}Cf(\eta_n)\to b\in\mathbb{R}^c$, then
$T_n\tl\chi_c^2(\delta)$, where $\delta=b^\top\Sigma_f^{-1}b$, $\chi_c^2(\delta)$ denotes the noncentral chi-square distribution with $c$ degrees of freedom and expectation $c+\delta$, and $\eta$ and  $\Sigma_f$ is built from the limits in Assumption \ref{as:B2}.
\end{enumerate}
\end{proposition}

Let  $\alpha \in (0,1)$. From Proposition \ref{prop:waldf}, the test that rejects $H_0$ when \begin{equation}
    T_n > \chi^2_{c, 1-\alpha},
    \label{eq:reg_crit}
\end{equation} where $\chi^2_{c, 1-\alpha}$ denotes the $(1-\alpha)$-quantile of the $\chi^2_{c}$ distribution, has asymptotic level $\alpha$. Proposition~\ref{prop:waldf} also characterizes the alternatives that the Wald test can detect. The test is consistent for fixed alternatives. Under local alternatives satisfying $N^{1/2}Cf(\eta_n)\to b$,
the statistic converges to a noncentral chi-square distribution, yielding a nontrivial asymptotic power $\beta(b) \geq \alpha$.

\begin{remark} \label{rem:TCC}
Note that Proposition~\ref{prop:waldf} also applies to the CC estimator, which is a particular case of AC. In fact, under CC analysis, $N^{(r)}=N^{(r,s)}=N_{full}$ for all $r,s$, and hence $\widehat\Omega_{rs}=
\widehat V^{CC}_{rs}/N_{full}$, where $\widehat V^{CC}$ denotes the jackknife estimator of Corollary~\ref{cor:jack} computed with all index sets equal to $D_{full}$, which is positive semidefinite by construction. Denote by $\widehat\Omega^{CC}$ and $\widehat\Omega_f^{CC}$ the resulting estimators of the covariance matrices of $\widehat\eta_{CC}$ and $f(\widehat\eta_{CC})$. In this case the matrix $M$ becomes $qJ_q$, where $J_q$ denotes the $q \times q $ matrix with all its entries equal to  1. From now on, we will denote by $T_n^{CC}$ the corresponding Wald statistic.
\end{remark}

\subsection{\texorpdfstring{$\boldsymbol{k}$}{k}-sample tests}\label{sec:application2}

The result of the previous subsection extends naturally to the comparison of parameters across several populations. Let $Z_{ij}=(X_{ij},M_{ij})$, $j=1,\ldots,n_i$, be $k$ independent samples of augmented observations, 
and let $f(\eta_i)\in\R^{d}$, $1\leq i\leq k$, where $\eta_i$ is the parameter vector of the $i$-th population, estimated by the AC vector $\widehat\eta_{AC,i}$ computed at the sample from that population, and $f$ is a fixed transformation with nonnull, continuous first-order partial derivatives at $\eta_i$, $1\leq i \leq k$. The hypothesis of interest is the equality of the transformed parameters,
\[
H_{0,k}:\ f(\eta_1)=\cdots=f(\eta_k) ,
\]
which can be written as $C f(\eta)=0$ for $f(\eta)=\big(f(\eta_1)^\top,\ldots,f(\eta_k)^\top\big)^\top$ and $C=\big[I_{k-1}\,,\,-\mathds{1}_{k-1}\big]\otimes I_{d}$, a full row rank matrix of dimension $(k-1)d\times kd$, where $I_d$ denotes the $d \times d$ identity matrix, $\mathds{1}_{k-1}$ denotes the $(k-1)$-dimensional vector of ones and $\otimes$ denotes the Kronecker product. Each sample carries its own estimator \eqref{eq:Omega}, $\widehat\Omega_i$, computed from its own availability counts $N_i^{(r)}$, $N_i^{(r,s)}$, and its own plug-in gradient, giving the $d\times d$ blocks

\[
\widehat\Omega_{f,i}=\nabla f\big(\widehat\eta_{AC,i}\big)\,\widehat\Omega_i\,\nabla f\big(\widehat\eta_{AC,i}\big)^\top ,
\qquad 1\leq i\leq k .
\]
Writing $f(\widehat\eta)=\big(f(\widehat\eta_{AC,1})^\top,\ldots,f(\widehat\eta_{AC,k})^\top\big)^\top$ and $\widehat\Omega_f=\diag\big(\widehat\Omega_{f,1},\ldots,\widehat\Omega_{f,k}\big)$, the block diagonal structure reflecting the independence of the samples, the statistic of the Wald test is
\begin{equation}\label{eq:ksample_stat}
T_{k,n}=\big\{Cf(\widehat\eta)\big\}^\top\Big(C\,\widehat\Omega_f\,C^\top\Big)^{-}\big\{Cf(\widehat\eta)\big\} .
\end{equation}

Within each population, Assumption~\ref{as:comparable} guarantees that the componentwise sample sizes grow at comparable rates. As is standard in $k$-sample problems, we also require the sample sizes across the different populations to be comparable. In our setting, this requirement is formalized in the following assumption.

\begin{assumption}\label{as:ksample}
Writing $N_i=n_i\sum_{r=1}^{q}\pi_i^{(r)}$ and $N=\sum_{i=1}^{k}N_i$, they satisfy
${N_i}/{N} \longrightarrow\ \lambda_i\in(0,1)$,
$1\leq i\leq k$.

\end{assumption}

\begin{proposition}\label{prop:ksample}
Suppose that $f:\mathbb{R}^q \to \mathbb{R}^d$ is continuous at $\eta_i$, has nonzero differential at $\eta_i$, $\nabla f(\eta_i)$, and each component of $f$ has continuous first-order partial derivatives in a neighborhood of $\eta_i$, $1\leq i \leq k$. Let $\Sigma_{f,i}=\nabla f(\eta_i)\big(V_i\had M_i\big)\nabla f(\eta_i)^\top$ and $\Sigma_f=C\diag\Big(\Sigma_{f,1}/\lambda_1,\ldots,\Sigma_{f,k}/\lambda_k\Big) C^{\top}$. Suppose that Assumption~\ref{as:ksample} holds, that $\Sigma_f$ is non-singular and that, for each population, either Assumption \ref{as:jack} holds or  $n_i\pi^{(r,s)}_{i}$ remains bounded, $1\leq r,s\leq q$. Let $T_{k,n}$ be defined as in \eqref{eq:ksample_stat}.
\begin{enumerate}
\item If Assumptions \ref{as:moments}--\ref{as:comparable} hold for each population and $Cf(\eta)=0$, then $T_{k,n}\tl\chi^2_{(k-1)d}$.
\item If Assumptions \ref{as:moments}--\ref{as:comparable} hold for each population and $Cf(\eta)\neq0$, then $T_{k,n}\tp\infty$.
\item  If, for each population, Assumptions \ref{as:B1}--\ref{as:B3} hold, with limits $\eta_i$ and $\Gamma_i$ in Assumption \ref{as:B2}, $\Gamma_i$ satisfies Assumption \ref{as:nondeg} and Assumptions \ref{as:avail}--\ref{as:comparable} hold, and if $N^{1/2}Cf(\eta_n)\to b\in\R^{(k-1)d}$, where $f(\eta_n)=\big(f(\eta_{1,n})^\top,\ldots,f(\eta_{k,n})^\top\big)^\top$ and $\eta_{i,n}$ is the parameter vector of the $i$-th population along the local sequence, then $T_{k,n}\tl\chi^2_{(k-1)d}(\delta)$, where $\delta=b^\top\Sigma_f^{-1}b$ and $\eta_i$ and $V_i=(m_rm_s\Gamma_{i,rs})_{1\le r,s\le q}$.
\end{enumerate}
\end{proposition}

Let  $\alpha \in (0,1)$. From Proposition \ref{prop:ksample}, the test that rejects $H_{0,k}$ when \begin{equation}
    T_{k,n} > \chi^2_{(k-1)d, 1-\alpha},
    \label{eq:reg_critk}
\end{equation}  has asymptotic level $\alpha$. Proposition~\ref{prop:ksample} also characterizes the alternatives that the Wald test can detect. The test is consistent for fixed alternatives. Under local alternatives satisfying $N^{1/2}Cf(\eta_n)\to b$,
the statistic converges to a noncentral chi-square distribution, yielding a nontrivial asymptotic power $\beta(b) \geq \alpha$.

\section{Comparison between AC and CC Wald tests}\label{sec:comparison}

The AC and CC Wald tests both have the correct asymptotic size and are consistent against any fixed alternative. Thus, we compare them along local alternatives.

Following Remark~\ref{rem:CC1}, the CC procedure is a special case of the AC with an expected sample size of $n\pi^{full}_n$. For the validity of Propositions \ref{prop:waldf} and \ref{prop:ksample}, Assumptions \ref{as:avail} and \ref{as:comparable} must be extended to this case.
\begin{assumption}\label{as:full}
$\pi^{full}_n\big/\sum_{s=1}^{q}\pi_n^{(s)}\to\kappa_{full} \geq 0 \text{ and } n\pi^{full}_n\to\infty.$
\end{assumption}



Under Assumptions~\ref{as:moments}--\ref{as:jack} and~\ref{as:full}, the conditions of Proposition~\ref{prop:waldf} are satisfied for both the AC and the CC Wald test statistics. Furthermore, from Remark~\ref{rem:TCC}, the asymptotic covariance matrix for the CC statistic reduces to $\Sigma_f^{CC} = q C \nabla f(\eta) V \nabla f(\eta)^{\top} C^{\top}$. Thus, both Wald statistics share the same asymptotic $\chi_c^2$ reference distribution, and their local powers are increasing functions of their respective noncentrality parameters. To compare them, we evaluate their performance along a common sequence of local alternatives satisfying $N^{1/2}Cf(\eta)\to b\in\R^c$. For the CC procedure, this sequence implies $(q N_{full})^{1/2}Cf(\eta)\to \big(q\kappa_{full}\big)^{1/2}b$. The resulting noncentrality parameters are formally compared in the following theorem. For a symmetric matrix $A$, we write $A\succeq 0$ to denote that $A$ is positive semidefinite.

\begin{theorem}\label{thm:comparison}
Suppose that the conditions of Proposition~\ref{prop:waldf} and Assumption~\ref{as:full} hold. Assume also that
$C\nabla f(\eta)V\nabla f(\eta)^\top C^\top$
is non-singular. Let $\delta_{AC}$ and $\delta_{CC}$ denote the noncentrality parameters corresponding to the AC and CC procedures under the common sequence of local alternatives $N^{1/2}Cf(\eta)\to b\in\R^c$, given by 
\begin{equation}\label{eq:noncentr}
\delta_{AC}=b^{\top}\Sigma_f^{-1}b,
\qquad
\delta_{CC}=\kappa_{full}\,b^{\top}\big\{C\,\nabla f(\eta)\,V\,\nabla f(\eta)^{\top}C^{\top}\big\}^{-1}b .
\end{equation}
Then,
\begin{enumerate}
\item If $\kappa_{full}=0$, then $\delta_{CC}=0<\delta_{AC}$ for every $b\in\R^{c}\setminus \{0\}$.
\item If $\kappa_{full}>0$, then $\delta_{AC}\geq\delta_{CC}$ for every $b\in\R^{c}$ if and only if  $C\nabla f(\eta)\big(V\had Q\big)\nabla f(\eta)^{\top}C^{\top}\succeq 0$, where $Q$ is the $q\times q$ matrix with entries $Q_{rs}=1\big/\kappa_{full}-M_{rs}$, $1\leq r,s\leq q$.
\end{enumerate}
\end{theorem}

Part 1 of Theorem~\ref{thm:comparison} establishes that, when $\kappa_{full}=0$, the AC test is strictly more powerful than the CC test under local alternatives in every nonzero direction. This is because, as $n$ increases, the expected number of complete observations becomes negligible relative to the expected number of observations available to the AC procedure. By contrast, when $\kappa_{full}>0$, the effective sample sizes of the two procedures are of the same order, and their relative local power depends on the condition given in part 2. However, this condition involves $V$, which is generally unknown in practice. We next examine two particular cases.

\begin{corollary}\label{cor:indep}
Suppose that the assumptions of Theorem~\ref{thm:comparison} hold and that $\Gamma_{rs}=0$ for all $r\neq s$. Then $\delta_{AC}\geq\delta_{CC}$ for every $b\in\R^{c}$.
\end{corollary}

Corollary~\ref{cor:indep} shows that the potential advantage of the CC procedure lies entirely in the off-diagonal entries of $V$. The next result evaluates a scenario where $V$ is proportional to an equicorrelation matrix, $V = v\{(1-\rho)I_q + \rho J_q\}$, where $J_q$ is the $\mathds{1}_q \mathds{1}_q^{\top}$. To ensure that $V$ is positive definite, the parameter $\rho$ must necessarily fall within the interval $\iota_q=(-1/(q-1), 1)$.

\begin{corollary}\label{cor:cs}
Suppose that the assumptions of Theorem~\ref{thm:comparison} hold with $\kappa_{full}>0$ and that $V=v\{(1-\rho)I_q+\rho J_q\}$, for some $v>0$ and $\rho\in \iota_q$. Let $D_Q=\diag(Q_{11},\ldots,Q_{qq})$, where $Q_{rs}$ are defined in Theorem \ref{thm:comparison},
\[
\mathcal R=\{\rho \in \iota_q:\ \delta_{AC}\ge\delta_{CC}\ \mbox{for every}\ b\in\R^c\},
\qquad
\mathcal R_b=\{\rho \in \iota_q :\ \delta_{AC}\ge\delta_{CC}\},\quad b\in\R^c\setminus\{0\},
\]
where $\delta_{AC}$ and $\delta_{CC}$ are defined in \eqref{eq:noncentr}. Then,
\begin{enumerate}
\item $\mathcal R$ is a relatively closed interval containing $0$ and, if $\Psi_D=C\nabla f(\eta)D_Q\nabla f(\eta)^\top C^\top$ is positive definite, $0$ is an interior point of $\mathcal R$.
\item If $C\nabla f(\eta)  \mathds{1}_q=0$, then, for every $b$, $\mathcal R_b$ is an interval of the form  
either $(-1/\{q-1\},\rho^*(b)]$ with $\rho^*(b)\ge0$, or the whole $\iota_q$.
\end{enumerate}
\end{corollary}
If $f$ is the identity function and $C$ is a contrast matrix, then part 2 of Corollary \ref{cor:cs} applies.

\begin{remark}\label{rem:kcomparison}
The comparison extends verbatim to the $k$-sample setting of Subsection~\ref{sec:application2}. Let $T_{k,n}^{CC}$ denote the statistic \eqref{eq:ksample_stat} with each block computed on the complete cases of its population, suppose that the conditions of Proposition~\ref{prop:ksample} hold, that Assumption~\ref{as:full} holds within each population, with limits $\kappa_{full,i}$, and that each $V_i$ is positive definite. If $\kappa_{full,i}>0$, $1\leq i\leq k$, then, along a common sequence of alternatives satisfying $N^{1/2}Cf\big(\eta\big)\to b\in\R^{(k-1)d}$, the noncentrality parameters are $\delta_{AC}=b^{\top}\Sigma_f^{-1}b$ and
\[
\delta_{CC}=b^{\top}\Big[C\diag\Big\{\big(\lambda_1\kappa_{full,1}\big)^{-1}\nabla f\big(\eta_1\big)V_1\nabla f\big(\eta_1\big)^{\top},\ldots,\big(\lambda_k\kappa_{full,k}\big)^{-1}\nabla f\big(\eta_k\big)V_k\nabla f\big(\eta_k\big)^{\top}\Big\}C^{\top}\Big]^{-1}b,
\]
and $\delta_{AC}\geq\delta_{CC}$ holds for every $b\in\R^{(k-1)d}$ if and only if
\[
C\diag\Big\{\lambda_1^{-1}\nabla f\big(\eta_1\big)\big(V_1\had Q_1\big)\nabla f\big(\eta_1\big)^{\top},\ldots,\lambda_k^{-1}\nabla f\big(\eta_k\big)\big(V_k\had Q_k\big)\nabla f\big(\eta_k\big)^{\top}\Big\}C^{\top}\succeq0,
\]
where $Q_i$ is the matrix of Theorem~\ref{thm:comparison} computed from the availability limits of the $i$-th population: the samples being independent, both statistics are self-normalized blockwise, and the argument of the proof of Theorem~\ref{thm:comparison} applies without change. In particular, the condition holds whenever $\nabla f\big(\eta_i\big)\big(V_i\had Q_i\big)\nabla f\big(\eta_i\big)^{\top}\succeq0$ for every population.
\end{remark}

In practice, it is not possible to determine in advance which of the two procedures is preferable, since the quantities needed to make such a decision are unknown. The next section proposes a procedure that overcomes these limitations.

\section{A maximum test}
\label{subsec:boot}

The results of the previous section show that neither the AC nor the
CC procedure is uniformly more powerful. Here we consider 
\(T_n^{\max}=\max(T_n,T_n^{CC})\).


{Notice that under $H_0$, we have that $C f(\eta)=0$, and hence we can write}
\[\begin{aligned}
    T_n&=\Big[C\left\{f\big(\widehat\eta_{AC}\big)-f\big(\eta\big)\right\}\Big]^\top
\Big(C\,\widehat\Omega_f C^\top\Big)^{-}
\Big[C\left\{f\big(\widehat\eta_{AC}\big)-f\big(\eta\big)\right\}\Big], \\
T^{CC}_n&=\Big[C\left\{f\big(\widehat\eta_{CC}\big)-f\big(\eta\big)\right\}\Big]^\top
\Big(C\,\widehat\Omega^{CC}_f C^\top\Big)^{-}
\Big[C\left\{f\big(\widehat\eta_{CC}\big)-f\big(\eta\big)\right\}\Big].
\end{aligned}\]
From the developments in the previous sections,
\[
\begin{aligned}
T_n&=\Big\{C\nabla f(\eta)\big(\widehat\eta_{AC}-\eta\big)\Big\}^\top
\Big(C\,\widehat\Omega_f C^\top\Big)^{-}
\Big\{C\nabla f(\eta)\big(\widehat\eta_{AC}-\eta\big)\Big\}+r_1 \\
&= \Big\{\,C\nabla f(\eta)\, \widehat{N}^{1/2} \widehat{\mathcal D}^{-1}_n \widehat{\mathcal D}_n\big(\widehat\eta_{AC}-\eta\big)\Big\}^\top
\Big(\widehat{N}\, C\,\widehat\Omega_f C^\top\Big)^{-} 
\Big\{C\nabla f(\eta)\, \widehat{N}^{1/2} \widehat{\mathcal D}^{-1}_n \widehat{\mathcal D}_n\big(\widehat\eta_{AC}-\eta\big)\Big\}+r_1 \\
&=\Big\{ \,C\nabla f(\eta)K\, \widehat{\mathcal D}_n\big(\widehat\eta_{AC}-\eta\big)\Big\}^\top
\Sigma_f^{-1}
\Big\{C\nabla f(\eta)K\, \widehat{\mathcal D}_n\big(\widehat\eta_{AC}-\eta\big)\Big\}+r_2,
\end{aligned}
\] 
where $r_2 \tp 0$, and $K$ is as in Corollary~\ref{cor:equal}. Analogously,
\[
\begin{aligned}
T^{CC}_n
&=\Big\{q^{1/2}\,C\nabla f(\eta)\,N_{full}^{1/2}\big(\widehat\eta_{CC}-\eta\big)\Big\}^\top
\Big(\Sigma_f^{CC}\Big)^{-1}
\Big\{q^{1/2}\,C\nabla f(\eta)\,N_{full}^{1/2}\big(\widehat\eta_{CC}-\eta\big)\Big\}+r_3,
\end{aligned}
\]
where $r_3 \tp 0$.  From the above expressions, it becomes clear that to derive the asymptotic null distribution of $T_n^{\max}$, we must first derive the joint asymptotic distribution of $\widehat\eta_{AC}-\eta$ and $\widehat\eta_{CC}-\eta$, which is given in the next result.

\begin{lemma}\label{cor:joint}
Suppose that Assumptions~\ref{as:moments}--\ref{as:comparable} and
\ref{as:full} hold. Then
\[
\Big\{
\widehat{\mathcal D}_n
\big(\widehat\eta_{AC}-\eta\big)^{\top},
N_{full}^{1/2}
\big(\widehat\eta_{CC}-\eta\big)^{\top}
\Big\}^{\top}
\tl
Z,
\]
where $Z=(Z_1^\top,Z_2^\top)^\top\sim\mathcal N_{2q}(0,\tilde V\had\tilde\varrho)$, $Z_1,Z_2\in\R^q$, and 
\[
\tilde V=
\begin{pmatrix}V&V\\V&V\end{pmatrix},
\qquad
\tilde\varrho=
\begin{pmatrix}\varrho&\Delta\\
\Delta^{\top}&J_q\end{pmatrix},
\]
and \(\Delta\) is the \(q\times q\) matrix with entries
\(\Delta_{rs}=(\kappa_{full}/\kappa_r)^{1/2}\),
\(1\leq r,s\leq q\).
\end{lemma}

Under $H_0$, and assuming that $\Sigma_f$ and $\Sigma_f^{CC}$ are non-singular, from Lemma~\ref{cor:joint} we have that $(T_n,T_n^{CC})\tl(\chi_A,\chi_C)$, where
\begin{equation}\label{eq:chis}
\chi_A=\big(C\nabla f(\eta)KZ_1\big)^\top\Sigma_f^{-1}\big(C\nabla f(\eta)KZ_1\big),
\quad
\chi_C=q\big(C\nabla f(\eta)Z_2\big)^\top\big(\Sigma_f^{CC}\big)^{-1}\big(C\nabla f(\eta)Z_2\big).
\end{equation}
 Since $\Sigma_f$ and $\Sigma_f^{CC}$ are the covariance matrices of $C\nabla f(\eta)KZ_1$ and $q^{1/2}C\nabla f(\eta)Z_2$, respectively, $\chi_A$ and $\chi_C$ are  generally dependent $\chi^2_c$ random variables. Define
\[
F_0(t)=\Prob\{\max(\chi_A,\chi_C)\leq t\},
\qquad t\in\mathbb R.
\]
Thus, under \(H_0\), \(T_n^{\max}\tl F_0\).
As \(F_0\) is unknown, we approximate the null distribution of $T_n^{\max}$ {using two methods: a bootstrap in the limit and a multiplier bootstrap}.

\begin{remark}\label{rem:joint_triangular}
If in Lemma~\ref{cor:joint} Assumptions~\ref{as:moments}--\ref{as:mcar} are replaced with Assumptions~\ref{as:B1}--\ref{as:B3}, with $\Gamma$ in Assumption~\ref{as:B2} satisfying Assumption~\ref{as:nondeg}, it can be easily checked that 
\[
\Big\{
\widehat{\mathcal D}_n
\big(\widehat\eta_{n,AC}-\eta_n\big)^{\top},
N_{full}^{1/2}
\big(\widehat\eta_{n,CC}-\eta_n\big)^{\top}
\Big\}^{\top}
\tl
Z,
\]
where $Z$ is as in Lemma~\ref{cor:joint}, with $V=(m_rm_s\Gamma_{rs})_{1\le r,s\le q}$.
\end{remark}

\subsection{Bootstrap in the limit}\label{sec:BIL}

The bootstrap in the limit consists in approximating the null distribution of $T_n^{\max}$ by means of an estimation of its asymptotic null distribution, $F_0$. Recall that $F_0$ depends on the unknown quantities $V$, $\varrho$, $\Delta$, $\nabla f(\eta)$, $K$, $\Sigma_f$ and $\Sigma_f^{CC}$. Thus, the bootstrap in the limit estimates $F_0$ by replacing each of them with a consistent estimator:
\begin{itemize} \itemsep=0pt
    \item  $\varrho$ is replaced with $\widehat\varrho$, defined in Section~\ref{sec:variance}, and $\Delta_{rs}$ is replaced with $(N_{full}/N^{(r)})^{1/2}$, $1\leq r,s\leq q$. Let $\widehat{\tilde \rho}$ denote the resulting estimator of $\tilde \varrho$.
    \item  $K$ is replaced with $\widehat K=\widehat N^{1/2}\widehat{\mathcal D}_n^{-1}$.
    \item $\nabla f(\eta)$ appears twice: once associated with the AC statistic and once with the CC statistic. Similarly, we have two estimators of $\eta$, corresponding to AC and CC. We replace $\eta$ accordingly.
    \item Both $\Sigma_f$ and $\Sigma^{CC}_f$ depend on $V$. As before, we have two estimators of $V$, corresponding to AC and CC. We replace $V$ accordingly, giving rise to the estimators $\widehat{NC}\widehat\Omega_fC^\top$ and $qN_{full}C\widehat\Omega_f^{CC}C^\top$ of $\Sigma_f$ and $\Sigma^{CC}_f$, respectively.
    \item To estimate $\tilde V$, we face the same problem as in the previous item, and we consider 
    $\widehat{\tilde V}=\Pi\,\Pi^\top$, with $\Pi=\{\widehat V_{+}^{1/2},(\widehat V^{CC})^{1/2}\}^\top$, where $A^{1/2}$ denotes the symmetric square root of $A$. This estimator of $\tilde V$ ensures that the resulting matrix is positive semidefinite.

\end{itemize}

Let $Z^{\dagger}=(Z_1^{\dagger\top},Z_2^{\dagger\top})^\top\sim\mathcal N_{2q}\big(0,\widehat{\tilde V}\had\widehat{\tilde\varrho}\big)$ be independent of the data. Define
\[
T_n^{\dagger}=\big(C\nabla f(\widehat\eta_{AC})\widehat KZ_1^{\dagger}\big)^\top\big(\widehat NC\widehat\Omega_fC^\top\big)^{-}\big(C\nabla f(\widehat\eta_{AC})\widehat KZ_1^{\dagger}\big),
\]
\[
T_n^{CC\dagger}=\big(q^{1/2}C\nabla f(\widehat\eta_{CC})Z_2^{\dagger}\big)^\top\big(qN_{full}C\widehat\Omega_f^{CC}C^\top\big)^{-}\big(q^{1/2}C\nabla f(\widehat\eta_{CC})Z_2^{\dagger}\big).
\]
For large $n$, the matrices $\widehat NC\widehat\Omega_fC^\top$ and $qN_{full}C\widehat\Omega_f^{CC}C^\top$ are non-singular and then, by construction,  the conditional distribution of each of  $T^{\dagger}$ and $T_n^{CC\dagger}$, given the data,  is $\chi^2_c$.

We set $T_n^{\max,\dagger}=\max(T_n^{\dagger},T_n^{CC\dagger})$ and approximate the null distribution of $T_n^{\max}$ by $F_n^{\dagger}(t)=\Prob\{T_n^{\max,\dagger}\leq t\mid\mathcal X_n\}$, where $\mathcal{X}_n=\{(X_j,M_j):1\leq j\leq n\}$. Given the data, $F_n^{\dagger}$ is the distribution of an explicit function of a $2q$-dimensional Gaussian vector, and hence it can be approximated by Monte Carlo simulation.

Let $\alpha \in (0,1)$ and let $c^{\dagger}_{1-\alpha}$ be the $(1-\alpha)$-quantile of $F^{\dagger}_n$. The properties of the test that rejects $H_0$ when $T^{\max}>c^{\dagger}_{1-\alpha}$ coincide with that of Theorem \ref{thm:boot} in the next subsection, as argued in Remark \ref{rem:consBIL}.

In the $k$-sample setting of Subsection~\ref{sec:application2}, since the samples are independent, the joint limiting distribution is that of $k$ independent vectors as in Lemma~\ref{cor:joint}, and the bootstrap in the limit generates them accordingly. Let $Z_i^{\dagger}=(Z_{i1}^{\dagger\top},Z_{i2}^{\dagger\top})^\top$, $1 \leq i \leq k,$ be independent $\mathcal N_{2q}\big(0,\widehat{\tilde V}_i\had\widehat{\tilde\varrho}_i\big)$ vectors, where $\widehat{\tilde V}_i$, $\widehat{\tilde\varrho}_i$, $\widehat{\mathcal D}_{n,i}$ and $N_{full,i}$ denote the quantities $\widehat{\tilde V}$, $\widehat{\tilde\varrho}$, $\widehat{\mathcal D}_{n}$ and $N_{full}$ computed from the $i$-th sample. The statistics $T_{k,n}^{\dagger}$ and $T_{k,n}^{CC\dagger}$ are defined analogously to the one-sample case, and we set $T_{k,n}^{\max,\dagger}=\max(T_{k,n}^{\dagger},T_{k,n}^{CC\dagger})$ and $F_{k,n}^{\dagger}(t)=\Prob\{T_{k,n}^{\max,\dagger}\leq t\mid\mathcal X_n\}$, where now $\mathcal X_n$ denotes the $k$ samples.

Similarly, the properties of the that rejects $H_{0,k}$ when $T_{k,n}^{\max}$ is grater than the upper $(1-\alpha)$-quantile of $F_{k,n}^{\dagger}$ coincide with that of Theorem \ref{prop:kboot} in the next subsection, as argued in Remark \ref{rem:kbil}.

\subsection{Multiplier bootstrap}\label{sec:MB}

Alternatively, to approximate the distribution of $T_n$ we consider the following multiplier bootstrap version,
\[
T_n^*
=
\Big[
C \big\{f(\widehat{\eta}_{AC}^{*})-f(\widehat{\eta}_{AC})\big\}\Big]^{\top}
\left(
C\widehat\Omega_f C^{\top}
\right)^{-}
\Big[
C \big\{f(\widehat{\eta}_{AC}^{*})-f(\widehat{\eta}_{AC})\big\}\Big],
\]
where 
\[\widehat\eta_{AC}^{*(r)}
=
\widehat{\eta}_{AC}^{(r)}+\frac{1}{N^{(r)}}\sum_{j=1}^n
W_j\widehat\psi_j^{(r)},\]
with (recall the definition of $D_r$ given in Section \ref{sec:variance})
\begin{equation}
    \widehat{\psi}_j^{(r)}
=
\begin{cases}
\left(N^{(r)}-1\right)
\left(
\widehat{\eta}_{AC}^{(r)}
-
\widehat{\eta}_{AC,-j}^{(r)}
\right), & j \in D_r,\\
0, & j \not \in D_r,
\end{cases}
\label{eq:psi_hat}
\end{equation}
 and  \(W_1,\ldots,W_n\) are i.i.d. random variables (the multipliers), independent of the data,
satisfying \(\E(W_1)=0\) and \(\V(W_1)=1\). Analogously, let 
\[
T_n^{CC*}
=
\left[
C \{f(\widehat{\eta}_{CC}^{*})-f(\widehat{\eta}_{CC})\}\right]^{\top}
\left(
C\widehat\Omega_f^{CC} C^{\top}
\right)^{-}
\left[
C \{f(\widehat{\eta}_{CC}^{*})-f(\widehat{\eta}_{CC})\}\right],
\]
where 
\[\widehat\eta_{CC}^{*(r)}
= \widehat{\eta}_{CC}^{(r)}+
\frac{1}{N_{full}}\sum_{j=1}^n
W_j\widehat\varphi_j^{(r)},\]
with
\begin{equation}
    \widehat\varphi_j^{(r)}
=
\begin{cases}
\left(N_{full}-1\right)
\left(
\widehat{\eta}_{CC}^{(r)}
-
\widehat{\eta}_{CC,-j}^{(r)}
\right), & j \in D_{full},\\
0, & j \not \in D_{full},
\end{cases}
\label{eq:phi_hat}
\end{equation}
and $D_{full} \subseteq \{1,\ldots, n\}$ denotes the set of indices corresponding to the observations used to estimate $\widehat{\eta}_{CC}$. Notice that the same multipliers are used for both $T_n^*$ and $T_n^{CC*}$. We then set \(T_n^{\max,*}=\max(T_n^*,T_n^{CC*})\). Let
\(F_n^*(t)=\Prob\{T_n^{\max,*}\leq t\mid\mathcal X_n\}\). The next theorem establishes that the conditional bootstrap
distribution, given $X_n$, consistently estimates \(F_0\).


\begin{theorem}\label{thm:boot}
Suppose that $f:\mathbb{R}^q \to \mathbb{R}^d$ is continuous at $\eta$, has nonzero differential at $\eta$, $\nabla f(\eta)$, and each component of $f$ has continuous first-order partial derivatives in a neighborhood of $\eta$. Let $\Sigma_f=C\,\nabla f(\eta)\big(V\had M\big)\nabla f(\eta)^\top C^\top$ and $\Sigma_f^{CC}=qC\,\nabla f(\eta)V\nabla f(\eta)^\top C^\top$, $C$ being a $c\times d$ matrix of full row rank, and suppose that $\Sigma_f$ and $\Sigma_f^{CC}$ are non-singular, that Assumption~\ref{as:jack} holds and that Assumption~\ref{as:full} holds with $\kappa_{full}>0$. Let $\alpha\in(0,1)$ and let $c^*_{1-\alpha}$ denote the $(1-\alpha)$-quantile of $F_n^*$.
\begin{enumerate}
\item If Assumptions \ref{as:moments}--\ref{as:comparable} hold and $Cf(\eta)=0$, then $\Prob\big(T_n^{\max}>c_{1-\alpha}^*\big)\to\alpha$.
\item If Assumptions \ref{as:moments}--\ref{as:comparable} hold and $Cf(\eta)\neq0$, then $\Prob\big(T_n^{\max}>c_{1-\alpha}^*\big)\to1$.
\item If Assumptions \ref{as:B1}--\ref{as:B3} hold, $\Gamma$ in Assumption \ref{as:B2} satisfies Assumption \ref{as:nondeg}, Assumptions \ref{as:avail}--\ref{as:comparable} hold, and $N^{1/2}Cf(\eta_n)\to b\in\mathbb{R}^c$, then $\Prob\big(T_n^{\max}>c_{1-\alpha}^*\big)\to\beta(b)$, where $\beta(b)\geq\alpha$, and $\eta$ and $V=(m_rm_s\Gamma_{rs})_{1\le r,s\le q}$ in $\Sigma_f$ and $\Sigma_f^{CC}$ are built from the limits in Assumption \ref{as:B2}.
\end{enumerate}
Moreover, in all three cases,
\begin{equation}
    \sup_{t\in\mathbb R}
\left|F_n^{*}(t)-F_0(t)\right|
\tp 0.
\label{eq:FnF0}
\end{equation}
\end{theorem}

From Theorem~\ref{thm:boot}, the test that
rejects \(H_0\) whenever \(T_n^{\max}>c_{1-\alpha}^*\) is asymptotically valid in the sense that its type-I error converges to $\alpha$, it is consistent against any fixed alternative and able to detect local alternatives satisfying $N^{1/2}Cf(\eta_n)\to b\in\R^c$.

\begin{remark}
\label{rem:BIL-MB}
    The multiplier bootstrap reweights the contributions of the observations with random multipliers, and it evaluates $f$ at the perturbed estimators. If $f(\widehat\eta^*_{AC})-f(\widehat\eta_{AC})$ and $f(\widehat\eta^*_{CC})-f(\widehat\eta_{CC})$ were replaced with their first-order approximations, $\nabla f(\widehat\eta_{AC})(\widehat\eta^*_{AC}-\widehat\eta_{AC})$ and $\nabla f(\widehat\eta_{CC})(\widehat\eta^*_{CC}-\widehat\eta_{CC})$, then, for standard normal multipliers, the resulting procedure would be a bootstrap in the limit with the empirical covariance matrix of the quantities defined in \eqref{eq:psi_hat} and \eqref{eq:phi_hat}  as the estimator of $\tilde V\had\tilde\varrho$. In this case, the conditional distribution given $\mathcal X_n$ of the vector $[\{\widehat{\mathcal D}_n(\widehat\eta^{*}_{AC}-\widehat \eta_{AC})\}^\top,\{N_{full}^{1/2}(\widehat\eta^{*}_{CC}-\widehat\eta_{CC} )\}^\top]^\top$ is exactly Gaussian with mean zero and covariance matrix converging to $\tilde V\had\tilde\varrho$ (see Lemma~\ref{lem:bootcov} in the Supporting Information).  Both approximations thus differ in the estimator of the covariance matrix  and in the terms of higher order of $f$. 
\end{remark}

\begin{remark}
\label{rem:consBIL}
   As noted in Remark \ref{rem:BIL-MB}, the statistics of the bootstrap in the limit are the plug-in version of the leading terms in the expansion of $T_n^*$ and $T_n^{CC*}$, and their consistency follows from the same arguments, with no remainder terms to control (see the proof of Theorem~\ref{thm:boot} in Section~\ref{sec:proofs} of the Supporting Information). Hence, Theorem~\ref{thm:boot}, and therefore the above properties, also hold when $F_n^*$ and $c^*_{1-\alpha}$ are replaced by $F_n^{\dagger}$, as defined in Subsection \ref{sec:BIL}, and its $(1-\alpha)$-quantile.
\end{remark}

The approach can be also applied to the \(k\)-sample setting. For each population
\(i=1,\ldots,k\), let \(\widehat\psi_{ij}^{(r)}\) and
\(\widehat\varphi_{ij}^{(r)}\) denote the  quantities
defined above for the AC and CC estimators, respectively. Let
\(\{W_{ij}:1\leq j\leq n_i,\ 1\leq i\leq k\}\) be independent
multipliers with \(\E(W_{ij})=0\) and \(\V(W_{ij})=1\), independent of
the data. The same multiplier is used for the AC and CC quantities
corresponding to the same observation. Define
\[
\widehat{\eta}_{AC,i}^{*(r)}
=
\widehat{\eta}_{AC,i}^{(r)}+\frac{1}{N_i^{(r)}}\sum_{j=1}^{n_i}
W_{ij}\widehat\psi_{ij}^{(r)},
\qquad
\widehat\eta_{CC,i}^{*(r)}
=
\widehat\eta_{CC,i}^{(r)}+\frac{1}{N_{full,i}}\sum_{j=1}^{n_i}
W_{ij}\widehat\varphi_{ij}^{(r)}.
\]
We stack the population-specific estimators $\widehat\eta_{AC}^{*}
=
\big(
\widehat\eta_{AC,1}^{*\top},\ldots,
\widehat\eta_{AC,k}^{*\top}
\big)^\top
$
and 
$
\widehat\eta_{CC}^{*}
=
\big(
\widehat\eta_{CC,1}^{*\top},\ldots,
\widehat\eta_{CC,k}^{*\top}
\big)^\top.$ The bootstrap Wald statistics are defined through the same
Wald forms as their one-sample counterparts. Specifically,
\[
T_{k,n}^{*}
=
\left[
C \{f(\widehat{\eta}_{AC}^{*})-f(\widehat{\eta}_{AC})\}\right]^{\top}
\left(
C\widehat\Omega_f C^{\top}
\right)^{-}
\left[
C \{f(\widehat{\eta}_{AC}^{*})-f(\widehat{\eta}_{AC})\}\right],
\]
and
\[
T_{k,n}^{CC*}
=
\left[
C \{f(\widehat{\eta}_{CC}^{*})-f(\widehat{\eta}_{CC})\}\right]^{\top}
\left(
C\widehat\Omega_f^{CC} C^{\top}
\right)^{-}
\left[
C \{f(\widehat{\eta}_{CC}^{*})-f(\widehat{\eta}_{CC})\}\right],
\]
where in this case $C$, $\widehat\Omega_f$ and $\widehat\Omega_f^{CC}$ are as defined in Subsection \ref{sec:application2}. We then set $T_{k,n}^{\max}=\max(T_{k,n},T_{k,n}^{CC})$, \(T_{k,n}^{\max,*}=\max(T_{k,n}^{*},T_{k,n}^{CC*})\), and let
\(F_{k,n}^{*}(t)=\Prob\{T_{k,n}^{\max,*}\leq t\mid\mathcal X_n\}\). Let $F_{0,k}$ denote the limiting null distribution function of $T_{k,n}^{\max}$. The following theorem shows that the weighted bootstrap consistently
estimates this null distribution.

\begin{theorem}\label{prop:kboot}
Suppose that the conditions on $f$ and $\Sigma_f$ of Proposition~\ref{prop:ksample} hold, that Assumption~\ref{as:ksample} holds, that $\Sigma_f^{CC}$, the matrix in Remark~\ref{rem:kcomparison}, is non-singular, and that Assumptions~\ref{as:jack} and~\ref{as:full} hold within each population, with $\kappa_{full,i}>0$, $1\leq i\leq k$. Let $\alpha\in(0,1)$ and let $c_{k,1-\alpha}^*$ denote the $(1-\alpha)$-quantile of $F_{k,n}^*$. 
\begin{enumerate}
\item If Assumptions \ref{as:moments}--\ref{as:comparable} hold for each population and $Cf(\eta)=0$, then $\Prob\big(T_{k,n}^{\max}>c_{k,1-\alpha}^*\big)\to\alpha$.
\item If Assumptions \ref{as:moments}--\ref{as:comparable} hold for each population and $Cf(\eta)\neq0$, then $\Prob\big(T_{k,n}^{\max}>c_{k,1-\alpha}^*\big)\to1$.
\item If, for each population, Assumptions \ref{as:B1}--\ref{as:B3} hold, with limits $\eta_i$ and $\Gamma_i$ in Assumption \ref{as:B2}, $\Gamma_i$ satisfies Assumption \ref{as:nondeg} and Assumptions \ref{as:avail}--\ref{as:comparable} hold, and if $N^{1/2}Cf(\eta_n)\to b\in\mathbb R^{(k-1)d}$, with $f(\eta_n)$ as in Proposition~\ref{prop:ksample}, then $\Prob\big(T_{k,n}^{\max}>c_{k,1-\alpha}^*\big)\to\beta_k(b)$, where $\beta_k(b)\geq\alpha$, and $\eta_i$ and $V_i=(m_rm_s\Gamma_{i,rs})_{1\le r,s\le q}$ in $\Sigma_f$ and $\Sigma_f^{CC}$ are built from the limits in Assumption \ref{as:B2}.
\end{enumerate}
Moreover, in all three cases,
\begin{equation}
    \sup_{t\in\mathbb R}
\left|F_{k,n}^{*}(t)-F_{0,k}(t)\right|
\tp 0.
\label{eq:FnF0_k}
\end{equation}
\end{theorem}

From Theorem~\ref{prop:kboot}, the test that
rejects \(H_{0,k}\) whenever \(T_{k,n}^{\max}>c_{k,1-\alpha}^*\)  is asymptotically valid in the sense that its type-I error converges to $\alpha$, it is consistent against any fixed alternative and able to detect local alternatives satisfying $N^{1/2}Cf(\eta_n)\to b\in\R^{(k-1)d}$.

\begin{remark}
\label{rem:kbil}
    As in the one-sample case, the statistics of the $k$-sample bootstrap in the limit are the plug-in version of the leading terms in the expansion of $T_{k,n}^{*}$ and $T_{k,n}^{CC*}$, so that Theorem~\ref{prop:kboot}, and therefore the above properties, also hold when $F_{k,n}^*$ and $c^*_{k,1-\alpha}$ are replaced by $F_{k,n}^{\dagger}$, as defined in Subsection \ref{sec:BIL}, and its $(1-\alpha)$-quantile.
\end{remark}

In practice, the conditional bootstrap distributions are approximated by
Monte Carlo simulation using $B$ independent multiplier draws, and those of the bootstrap in the limit using $B$ independent Gaussian draws.

    \section{Numerical studies}\label{sec:applications}

   The results in the previous sections are asymptotic. This section examines the finite sample performance of the proposed tests for two problems: testing the equality of the variances of the components of a random vector, a linear hypothesis on $U$-statistics of degree $2$ of the form considered in Section~\ref{sec:onesample}, and comparing the coefficients of variation (CV) of $k$ populations, a hypothesis on smooth functions of $U$-statistics of degrees $1$ and $2$ addressed with the $k$-sample tests of Subsection~\ref{sec:application2}. 
    All simulated rejection frequencies are estimated with $10^4$ Monte Carlo replicates. The critical values of the bootstrap in the limit (BL) and the multiplier bootstrap (MB) are approximated with $B=10^{3}$ draws in both cases. For each example, we first use simulated data and then consider data from the 2017--2018 National Health and Nutrition Examination Survey (NHANES).

    \subsection{Equality of variances: simulation study}\label{sec:application1}\label{subsect:var.setting}\label{sec:app:vardim}
    Let $X= (X^{(1)},\ldots,X^{(q)})^\top$ be a random vector with component variances $\sigma_{11},\ldots,\sigma_{qq}$, and consider the hypothesis
    \begin{equation}
        H_0^{(q)}:\ \sigma_{11}=\cdots=\sigma_{qq},
        \label{eq:H0_app1}
    \end{equation}
    which can be expressed as $H_0^{(q)}:\ C\eta=0$, where $\eta=(\sigma_{11},\ldots,\sigma_{qq})^\top$ and
    \[
    C=
    \begin{pmatrix}
    1 & -1 & 0 & \cdots & 0\\
    0 & 1 & -1 & \ddots & \vdots\\
    \vdots & \ddots & \ddots & \ddots & 0\\
    0 & \cdots & 0 & 1 & -1
    \end{pmatrix}
    \in\mathbb{R}^{(q-1)\times q}.
    \]
     The components of $\eta$ are estimated by  a degree-2 $U$-statistics with kernels $h^{(r)}(x_1,x_2)=(x_1^{(r)}-x_2^{(r)})^2/2$. Let $\varpi^{(\ell)}=\Prob(M^{(\ell)}_j =1)$ and $\varpi=(\varpi^{(1)}, \ldots, \varpi^{(r)})$. Note that in this application, since $A_r=\{r\}$, it follows that  $\pi^{(r)}=\varpi^{(r)}$.

    We consider the above setting with $X \sim \mathcal{N}_q(0,\Sigma)$, where $\Sigma=(1-\rho_x)I_q+\rho_x J_q$. In this setting, from part 2 of Corollary~\ref{cor:cs}, we obtain that $\delta_{AC}\geq \delta_{CC}$ whenever $\rho_x \leq \rho_x^*$, with equality $\delta_{AC} = \delta_{CC}$ holding if and only if $\rho_x = \rho_x^*$. Here, the threshold $\rho_x^*$ depends on both the missingness mechanism and the alternative under consideration. Five missingness mechanisms are considered:
    \begin{enumerate} \itemsep=0pt
    \item [Case 1.] Complete data with $\varpi=(1,1,1,1)$.
    \item [Case 2.] Independent  with
          $\varpi
          =(0.75,0.75,0.75,0.75)$.
    \item [Case 3.] Independent  with
          $\varpi
          =(0.9,0.8,0.7,0.6)$.
    \item [Case 4.] Dependent generated through a Gaussian copula   
          $\theta=0.3$ and marginal probabilities as in Case 2.
    
    \item [Case 5.] Dependent  generated through a Gaussian copula mechanism  with
          $\theta=-0.3$ and marginal probabilities as in Case 2.
    \end{enumerate}
    See Section~\ref{sec:copula} of the Supporting Information for a detailed description of the Gaussian copula missingness mechanism. Table~\ref{tab:app:choice} collects the values of $\rho^*_x$ of the six experiments. 
    \begin{table}[t]
    \centering
    \caption{Values of $\rho_x^{*}$ at which the CC equals the local asymptotic power of the AC. Below $\rho_x^{*}$, the AC is asymptotically preferable; above it, the CC.}
    \label{tab:app:choice}
    \begin{tabular}{lccc}
    \toprule
    Missingness mechanism & $q=2$ & $q=3$ & $q=4$ \\
    \midrule
    Indep.\ $\varpi=(.75,.75,.75,.75)$       & $0.756$ & $0.870$ & $0.920$ \\
    Indep.\ $\varpi=(.9,.8,.7,.6)$ & $0.732$ & $0.869$ & $0.942$ \\
    Gauss.\ cop.\ $\theta=0.3$     & $0.754$ & $0.866$ & $0.914$ \\
    Gauss.\ cop.\ $\theta=-0.3$    & $0.758$ & $0.874$ & $0.925$ \\
    \bottomrule
    \end{tabular}
    \end{table}
    It is worth noting that the results in Table~\ref{tab:app:choice} pertain specifically to the considered setting. Altering the distribution of the vector $X$, the missingness probabilities, or the underlying hypotheses may lead to substantially different results.
    For each generated dataset the null hypothesis \eqref{eq:H0_app1} is tested with the tests AC (using the available data), CC (using the complete data) and the test based on the statistic $T^{\max}_n$, with two approximations of its null distribution: BL and MB.

    Table~\ref{tab:app1_level} reports the empirical sizes for $q=2,3,4$ and $n\in\{50,100,250,500\}$, at the nominal level $\alpha=0.05$. Table~\ref{tab:app1_sup_level} complements it with the average sample sizes actually used by each approach.
    Looking at Table~\ref{tab:app1_level} we observe that for complete data and small sample sizes, the empirical size increases with $q$. The same phenomenon is observed all tests when there are missing data, although the discrepancy from the nominal level is more pronounced because the effective sample size is substantially smaller (see Table~\ref{tab:app1_sup_level}). As $n$ increases, the empirical sizes of all methods approach the nominal level.
    
   \begin{table}[h]
    \centering
    \caption{Empirical sizes (in \%) at the nominal level $\alpha=0.05$.}
    \label{tab:app1_level}
    \resizebox{\textwidth}{!}{
    \begin{tabular}{ll|ccc|ccc|ccc|ccc|ccc|ccc|ccc|ccc}
    \toprule
    \multirow{3}{*}{Case} &
     &
    \multicolumn{12}{c|}{$\rho_x=0.3$} &
    \multicolumn{12}{c}{$\rho_x=0.95$} \\
    \cmidrule(lr){3-14}\cmidrule(lr){15-26}
    &
    &
    \multicolumn{3}{c|}{AC} &
    \multicolumn{3}{c|}{CC} &
    \multicolumn{3}{c|}{BL} &
    \multicolumn{3}{c|}{MB} &
    \multicolumn{3}{c|}{AC} &
    \multicolumn{3}{c|}{CC} &
    \multicolumn{3}{c|}{BL} &
    \multicolumn{3}{c}{MB} \\
    \cmidrule(lr){3-5}\cmidrule(lr){6-8}\cmidrule(lr){9-11}\cmidrule(lr){12-14}
    \cmidrule(lr){15-17}\cmidrule(lr){18-20}\cmidrule(lr){21-23}\cmidrule(lr){24-26}
    &$n\backslash q$&
    $2$ & $3$ & $4$ &
    $2$ & $3$ & $4$ &
    $2$ & $3$ & $4$ &
    $2$ & $3$ & $4$ &
    $2$ & $3$ & $4$ &
    $2$ & $3$ & $4$ &
    $2$ & $3$ & $4$ &
    $2$ & $3$ & $4$ \\
    \midrule
    \multirow{4}{*}{1.}
    &  50 & 5.0 & 6.3 & 7.5 & 5.0 & 6.3 & 7.5 & 4.8 & 6.0 & 7.5 & 5.2 & 6.4 & 8.1 & 5.0 & 5.0 & 5.8 & 5.0 & 5.0 & 5.8 & 5.6 & 5.7 & 5.9 & 6.0 & 6.2 & 6.4 \\
    & 100 & 5.5 & 5.9 & 6.2 & 5.5 & 5.9 & 6.2 & 5.3 & 5.7 & 6.3 & 5.4 & 5.8 & 6.5 & 4.8 & 5.1 & 5.2 & 4.8 & 5.1 & 5.2 & 4.9 & 5.2 & 5.5 & 5.0 & 5.4 & 5.8 \\
    & 250 & 5.3 & 5.8 & 5.5 & 5.3 & 5.8 & 5.5 & 5.0 & 5.5 & 5.6 & 5.1 & 5.6 & 5.6 & 5.0 & 5.1 & 5.0 & 5.0 & 5.1 & 5.0 & 4.7 & 5.1 & 5.5 & 4.7 & 5.2 & 5.6 \\
    & 500 & 5.2 & 5.0 & 5.2 & 5.2 & 5.0 & 5.2 & 5.2 & 5.2 & 5.4 & 5.2 & 5.1 & 5.5 & 4.8 & 4.7 & 4.9 & 4.8 & 4.7 & 4.9 & 5.0 & 5.1 & 5.2 & 5.1 & 5.0 & 5.1 \\
    \midrule
    \multirow{4}{*}{2.}
    &  50 & 5.2 & 6.7 & 8.3 & 5.0 & 7.3 & 12.4 & 5.0 & 7.5 & 13.1 & 5.4 & 8.4 & 14.4 & 3.4 & 2.4 & 1.8 & 4.4 & 5.7 & 9.1 & 4.2 & 4.2 & 5.9 & 4.7 & 5.4 & 8.2 \\
    & 100 & 4.8 & 5.4 & 6.4 & 4.7 & 6.0 & 8.9 & 4.5 & 6.5 & 9.2 & 4.8 & 7.0 & 9.6 & 4.1 & 2.9 & 2.2 & 5.1 & 5.4 & 6.7 & 4.6 & 4.0 & 4.7 & 5.1 & 5.0 & 6.1 \\
    & 250 & 5.0 & 5.1 & 5.6 & 5.0 & 5.7 & 6.6 & 5.0 & 5.3 & 6.6 & 5.0 & 5.6 & 6.9 & 4.7 & 3.9 & 3.3 & 5.3 & 5.1 & 5.5 & 4.6 & 4.3 & 4.2 & 4.8 & 4.8 & 4.8 \\
    & 500 & 4.9 & 5.1 & 5.4 & 5.0 & 5.5 & 6.1 & 5.3 & 5.5 & 6.1 & 5.2 & 5.6 & 6.1 & 4.7 & 4.6 & 4.4 & 4.8 & 4.5 & 5.1 & 4.7 & 4.7 & 4.4 & 4.8 & 4.9 & 4.9 \\
    \midrule
    \multirow{4}{*}{3.}
    &  50 & 5.4 & 6.9 & 7.8 & 5.2 & 7.5 & 13.8 & 5.2 & 7.6 & 14.5 & 5.4 & 8.3 & 15.9 & 3.6 & 2.7 & 1.9 & 5.0 & 5.7 & 9.7 & 4.3 & 4.5 & 6.3 & 4.9 & 5.4 & 8.9 \\
    & 100 & 5.1 & 5.9 & 7.0 & 5.1 & 6.1 & 9.4 & 5.0 & 6.8 & 9.4 & 5.4 & 7.0 & 9.9 & 4.2 & 3.2 & 2.0 & 4.8 & 5.5 & 6.8 & 4.1 & 4.1 & 4.5 & 4.3 & 4.8 & 5.8 \\
    & 250 & 4.8 & 5.1 & 5.4 & 5.1 & 5.5 & 6.9 & 5.0 & 5.5 & 6.6 & 5.1 & 5.8 & 7.1 & 4.9 & 3.8 & 3.4 & 5.0 & 5.3 & 6.2 & 5.1 & 5.0 & 4.5 & 5.2 & 5.4 & 5.2 \\
    & 500 & 5.1 & 5.2 & 5.9 & 5.0 & 5.1 & 6.0 & 5.3 & 5.6 & 6.0 & 5.4 & 5.7 & 6.2 & 4.8 & 4.7 & 4.0 & 5.3 & 5.1 & 5.6 & 5.1 & 4.9 & 4.8 & 5.2 & 5.2 & 5.2 \\
    \midrule
    \multirow{4}{*}{4.}
    &  50 & 5.1 & 6.5 & 8.1 & 4.8 & 7.9 & 13.0 & 5.3 & 8.0 & 12.9 & 5.8 & 8.8 & 14.4 & 3.5 & 2.5 & 1.8 & 4.8 & 5.7 & 8.3 & 4.1 & 4.3 & 5.6 & 4.9 & 5.4 & 8.0 \\
    & 100 & 4.9 & 5.7 & 6.7 & 4.9 & 6.3 & 8.3 & 5.3 & 6.6 & 9.2 & 5.6 & 6.8 & 9.7 & 4.0 & 3.1 & 2.4 & 5.2 & 5.3 & 5.9 & 4.4 & 4.3 & 4.8 & 5.0 & 4.8 & 6.1 \\
    & 250 & 4.5 & 4.8 & 5.5 & 4.9 & 5.2 & 6.2 & 5.1 & 5.5 & 6.4 & 5.1 & 5.7 & 6.4 & 4.9 & 4.3 & 3.6 & 5.2 & 5.1 & 5.3 & 4.7 & 4.5 & 4.5 & 5.0 & 4.9 & 5.2 \\
    & 500 & 4.7 & 5.1 & 5.0 & 5.2 & 5.3 & 5.7 & 4.8 & 5.3 & 5.7 & 4.8 & 5.3 & 5.8 & 4.9 & 4.3 & 4.1 & 5.1 & 5.4 & 5.4 & 5.0 & 4.9 & 4.7 & 5.2 & 5.3 & 5.2 \\
    \midrule
    \multirow{4}{*}{5.}
    &  50 & 5.0 & 6.3 & 7.9 & 4.9 & 8.3 & 14.2 & 5.0 & 8.1 & 14.2 & 5.5 & 8.9 & 15.7 & 3.5 & 2.2 & 1.8 & 5.5 & 6.0 & 9.3 & 3.9 & 4.3 & 6.3 & 4.7 & 5.5 & 8.5 \\
    & 100 & 5.1 & 6.0 & 6.8 & 5.1 & 6.8 & 9.7 & 5.3 & 6.4 & 9.3 & 5.4 & 6.8 & 9.9 & 4.0 & 2.7 & 2.3 & 4.7 & 5.7 & 6.4 & 4.3 & 3.9 & 4.4 & 4.8 & 4.8 & 6.0 \\
    & 250 & 5.0 & 5.2 & 5.7 & 5.2 & 5.6 & 7.0 & 5.1 & 5.6 & 6.8 & 5.4 & 5.8 & 7.1 & 4.6 & 3.8 & 3.3 & 5.3 & 5.3 & 5.4 & 4.7 & 4.0 & 4.3 & 4.9 & 4.5 & 5.0 \\
    & 500 & 4.9 & 4.8 & 4.9 & 5.1 & 5.3 & 5.9 & 5.1 & 5.4 & 5.8 & 5.0 & 5.5 & 5.8 & 4.8 & 4.5 & 4.2 & 5.0 & 5.2 & 5.1 & 5.2 & 4.5 & 4.5 & 5.3 & 4.8 & 4.9 \\
    \bottomrule
    \end{tabular}
    }
\end{table}

    \begin{table}[h]
    \centering
    \caption{Average sample sizes in the size study of Table~\ref{tab:app1_level}. For each $q$,
    $\bar N_{AC}$ is the average of the mean per-component available sample sizes among the $q$ components involved, and $\bar N_{CC}$ is the average CC sample size.}
    \label{tab:app1_sup_level}
    
    \begin{tabular}{ll|cc|cc|cc|cc|cc|cc}
    \toprule
    \multirow{3}{*}{Case} &
    \multirow{3}{*}{$n$} &
    \multicolumn{6}{c|}{$\rho_x=0.3$} &
    \multicolumn{6}{c}{$\rho_x=0.95$} \\
    \cmidrule(lr){3-8}\cmidrule(lr){9-14}
    &
    &
    \multicolumn{2}{c|}{$q=2$} &
    \multicolumn{2}{c|}{$q=3$} &
    \multicolumn{2}{c|}{$q=4$} &
    \multicolumn{2}{c|}{$q=2$} &
    \multicolumn{2}{c|}{$q=3$} &
    \multicolumn{2}{c}{$q=4$} \\
    \cmidrule(lr){3-4}\cmidrule(lr){5-6}\cmidrule(lr){7-8}
    \cmidrule(lr){9-10}\cmidrule(lr){11-12}\cmidrule(lr){13-14}
    &&
    $\bar N_{AC}$ & $\bar N_{CC}$ &
    $\bar N_{AC}$ & $\bar N_{CC}$ &
    $\bar N_{AC}$ & $\bar N_{CC}$ &
    $\bar N_{AC}$ & $\bar N_{CC}$ &
    $\bar N_{AC}$ & $\bar N_{CC}$ &
    $\bar N_{AC}$ & $\bar N_{CC}$ \\
    \midrule
    \multirow{4}{*}{1.}
    & 50
    & 50 & 50 & 50 & 50 & 50 & 50
    & 50 & 50 & 50 & 50 & 50 & 50 \\
    & 100
    & 100 & 100 & 100 & 100 & 100 & 100
    & 100 & 100 & 100 & 100 & 100 & 100 \\
    & 250
    & 250 & 250 & 250 & 250 & 250 & 250
    & 250 & 250 & 250 & 250 & 250 & 250 \\
    & 500
    & 500 & 500 & 500 & 500 & 500 & 500
    & 500 & 500 & 500 & 500 & 500 & 500 \\
    \midrule
    \multirow{4}{*}{2.}
    & 50
    & 38 & 28 & 37 & 21 & 38 & 16
    & 38 & 28 & 38 & 21 & 38 & 16 \\
    & 100
    & 75 & 56 & 75 & 42 & 75 & 32
    & 75 & 56 & 75 & 42 & 75 & 32 \\
    & 250
    & 188 & 141 & 188 & 106 & 188 & 79
    & 187 & 141 & 188 & 106 & 188 & 79 \\
    & 500
    & 375 & 281 & 375 & 211 & 375 & 158
    & 375 & 281 & 375 & 211 & 375 & 158 \\
    \midrule
    \multirow{4}{*}{3.}
    & 50
    & 42 & 36 & 40 & 25 & 38 & 15
    & 42 & 36 & 40 & 25 & 37 & 15 \\
    & 100
    & 85 & 72 & 80 & 50 & 75 & 30
    & 85 & 72 & 80 & 50 & 75 & 30 \\
    & 250
    & 213 & 180 & 200 & 126 & 187 & 76
    & 212 & 180 & 200 & 126 & 187 & 75 \\
    & 500
    & 425 & 360 & 400 & 252 & 375 & 151
    & 425 & 360 & 400 & 252 & 375 & 151 \\
    \midrule
    \multirow{4}{*}{4.}
    & 50
    & 38 & 28 & 38 & 22 & 37 & 17
    & 38 & 28 & 38 & 22 & 38 & 17 \\
    & 100
    & 75 & 57 & 75 & 43 & 75 & 34
    & 75 & 57 & 75 & 44 & 75 & 34 \\
    & 250
    & 188 & 142 & 188 & 109 & 188 & 84
    & 187 & 142 & 187 & 109 & 187 & 84 \\
    & 500
    & 375 & 284 & 375 & 218 & 375 & 168
    & 375 & 284 & 375 & 218 & 375 & 168 \\
    \midrule
    \multirow{4}{*}{5.}
    & 50
    & 38 & 28 & 38 & 20 & 38 & 15
    & 38 & 28 & 38 & 20 & 38 & 15 \\
    & 100
    & 75 & 56 & 75 & 41 & 75 & 30
    & 75 & 56 & 75 & 41 & 75 & 30 \\
    & 250
    & 188 & 139 & 188 & 102 & 187 & 74
    & 188 & 139 & 187 & 102 & 187 & 74 \\
    & 500
    & 375 & 278 & 375 & 204 & 375 & 148
    & 375 & 278 & 375 & 204 & 375 & 148 \\
    \bottomrule
    \end{tabular}
    \end{table}

    For the power study we set $\sigma_{22}=1+0.2$, keeping the remaining variances at $1$ and the correlations at $\rho_x$. Table~\ref{tab:app1_powadj} reports size-adjusted versions of the empirical powers, obtained by the probit method of \cite{lloyd2005}: if $\widehat\beta$ denotes the raw power of a test and $\widehat\alpha$ its empirical size in the matched null experiment of Table~\ref{tab:app1_level}, the size-adjusted power at the nominal level $\alpha^*$ is
    \begin{equation}\label{eq:app1_lloyd}
    \widehat R(\alpha^*)=\Phi\big\{\Phi^{-1}(\widehat\beta)-\Phi^{-1}(\widehat\alpha)+\Phi^{-1}(\alpha^*)\big\},
    \end{equation}
    which estimates the power the test would have if its size were exactly $\alpha^*$.  
    
    Looking at Table~\ref{tab:app1_powadj}, we observe that for $\rho_x=0.3$ the AC method outperforms the CC method in all the configurations considered, as expected from the results in Table~\ref{tab:app:choice}. For $\rho_x=0.95$, the results in Table~\ref{tab:app:choice} predict that, the CC method has higher asymptotic local power than the AC method. Table~\ref{tab:app1_powadj} shows that, for $q=4$ and $n=50,100$, the AC method still exhibits higher power. However, in general as $n$ increases, the CC method becomes more powerful, in agreement with the asymptotic theory. Note that, in almost all cases, both BL and MB lie between AC and CC, generally closer to the method with better performance, and their size-adjusted powers are very close.

 \begin{table}[h]
    \centering
    \caption{Size-adjusted powers (in \%), at the nominal level $\alpha=0.05$.}
    \label{tab:app1_powadj}
    \resizebox{\textwidth}{!}{
    \begin{tabular}{ll|ccc|ccc|ccc|ccc|ccc|ccc|ccc|ccc}
    \toprule
    \multirow{3}{*}{Case} &
     &
    \multicolumn{12}{c|}{$\rho_x=0.3$} &
    \multicolumn{12}{c}{$\rho_x=0.95$} \\
    \cmidrule(lr){3-14}\cmidrule(lr){15-26}
    &
    &
    \multicolumn{3}{c|}{AC} &
    \multicolumn{3}{c|}{CC} &
    \multicolumn{3}{c|}{BL} &
    \multicolumn{3}{c|}{MB} &
    \multicolumn{3}{c|}{AC} &
    \multicolumn{3}{c|}{CC} &
    \multicolumn{3}{c|}{BL} &
    \multicolumn{3}{c}{MB} \\
    \cmidrule(lr){3-5}\cmidrule(lr){6-8}\cmidrule(lr){9-11}\cmidrule(lr){12-14}
    \cmidrule(lr){15-17}\cmidrule(lr){18-20}\cmidrule(lr){21-23}\cmidrule(lr){24-26}
    &$n\backslash q$&
    $2$ & $3$ & $4$ &
    $2$ & $3$ & $4$ &
    $2$ & $3$ & $4$ &
    $2$ & $3$ & $4$ &
    $2$ & $3$ & $4$ &
    $2$ & $3$ & $4$ &
    $2$ & $3$ & $4$ &
    $2$ & $3$ & $4$ \\
    \midrule
    \multirow{4}{*}{1.}
    &  50 & 9.7 & 8.3 & 8.0 & 9.7 & 8.3 & 8.0 & 10.2 & 8.5 & 7.9 & 10.0 & 8.6 & 7.8 & 49.1 & 46.3 & 40.5 & 49.1 & 46.3 & 40.5 & 45.9 & 43.2 & 39.7 & 45.8 & 43.0 & 39.5 \\
    & 100 & 14.0 & 12.5 & 11.7 & 14.0 & 12.5 & 11.7 & 14.5 & 13.5 & 11.8 & 14.6 & 13.7 & 11.9 & 80.9 & 80.7 & 77.7 & 80.9 & 80.7 & 77.7 & 79.5 & 80.3 & 77.1 & 79.7 & 80.3 & 76.9 \\
    & 250 & 31.3 & 28.4 & 27.8 & 31.3 & 28.4 & 27.8 & 32.2 & 29.7 & 27.2 & 31.9 & 29.5 & 27.7 & 99.4 & 99.7 & 99.8 & 99.4 & 99.7 & 99.8 & 99.6 & 99.8 & 99.7 & 99.6 & 99.8 & 99.7 \\
    & 500 & 56.3 & 57.3 & 54.0 & 56.3 & 57.3 & 54.0 & 56.1 & 56.5 & 53.4 & 56.2 & 57.3 & 53.0 & 100 & 100 & 100 & 100 & 100 & 100 & 100 & 100 & 100 & 100 & 100 & 100 \\
    \midrule
    \multirow{4}{*}{2.}
    &  50 & 7.8 & 6.9 & 6.7 & 7.0 & 6.5 & 6.0 & 8.2 & 7.3 & 6.3 & 8.3 & 7.2 & 6.4 & 17.1 & 16.4 & 15.0 & 29.5 & 17.0 & 9.7 & 24.9 & 16.6 & 10.7 & 26.1 & 17.0 & 11.1 \\
    & 100 & 12.6 & 11.1 & 9.7 & 10.7 & 8.4 & 6.7 & 12.2 & 9.8 & 8.0 & 12.0 & 9.6 & 8.2 & 29.1 & 29.4 & 27.4 & 52.8 & 36.5 & 21.3 & 48.7 & 37.8 & 24.7 & 48.4 & 37.5 & 25.3 \\
    & 250 & 24.7 & 23.0 & 20.6 & 19.7 & 14.0 & 10.4 & 23.2 & 20.3 & 15.8 & 23.6 & 19.9 & 15.6 & 58.2 & 59.9 & 58.8 & 91.1 & 81.9 & 62.8 & 89.9 & 83.6 & 71.4 & 89.9 & 83.3 & 71.9 \\
    & 500 & 44.1 & 43.6 & 40.6 & 34.7 & 25.7 & 17.4 & 40.9 & 37.9 & 33.1 & 41.6 & 37.7 & 33.4 & 87.1 & 87.9 & 86.9 & 99.8 & 99.1 & 95.0 & 99.6 & 99.0 & 96.8 & 99.6 & 98.9 & 96.6 \\
    \midrule
    \multirow{4}{*}{3.}
    &  50 & 8.0 & 7.2 & 7.6 & 8.0 & 6.1 & 5.4 & 7.9 & 6.9 & 6.1 & 8.2 & 6.9 & 6.1 & 24.7 & 20.3 & 17.3 & 35.2 & 20.5 & 9.0 & 32.8 & 20.1 & 11.1 & 33.1 & 21.4 & 11.6 \\
    & 100 & 13.0 & 11.4 & 9.9 & 12.5 & 8.8 & 6.5 & 12.5 & 9.2 & 8.3 & 12.3 & 9.6 & 8.6 & 41.6 & 37.8 & 37.0 & 65.0 & 44.0 & 20.8 & 63.6 & 46.2 & 27.9 & 63.9 & 46.3 & 28.4 \\
    & 250 & 28.5 & 25.4 & 22.3 & 25.2 & 16.2 & 9.6 & 26.3 & 21.4 & 16.3 & 26.6 & 20.9 & 16.0 & 74.4 & 71.9 & 68.8 & 96.5 & 88.7 & 59.0 & 95.5 & 89.1 & 73.2 & 95.4 & 88.8 & 73.1 \\
    & 500 & 48.4 & 45.6 & 40.6 & 44.2 & 30.4 & 16.9 & 46.0 & 40.1 & 34.4 & 46.2 & 40.3 & 34.4 & 95.8 & 93.9 & 93.4 & 99.9 & 99.8 & 93.0 & 100 & 99.8 & 97.7 & 100 & 99.8 & 97.5 \\
    \midrule
    \multirow{4}{*}{4.}
    &  50 & 8.1 & 7.0 & 7.1 & 7.5 & 6.0 & 5.8 & 8.1 & 6.8 & 6.3 & 7.9 & 6.8 & 6.4 & 18.2 & 16.8 & 15.9 & 28.8 & 17.8 & 11.0 & 25.6 & 17.0 & 11.6 & 25.8 & 18.1 & 11.6 \\
    & 100 & 12.6 & 11.3 & 9.8 & 10.8 & 8.1 & 7.3 & 11.3 & 9.2 & 7.8 & 11.1 & 9.4 & 8.0 & 30.8 & 29.7 & 27.8 & 52.6 & 38.9 & 25.7 & 51.0 & 38.5 & 27.0 & 50.7 & 40.2 & 27.5 \\
    & 250 & 26.5 & 24.3 & 21.0 & 20.4 & 15.1 & 11.2 & 23.7 & 19.4 & 16.5 & 23.9 & 19.3 & 17.0 & 59.3 & 60.5 & 59.9 & 91.6 & 84.0 & 68.0 & 89.9 & 83.8 & 72.7 & 89.7 & 83.7 & 72.8 \\
    & 500 & 45.4 & 43.9 & 41.6 & 34.9 & 26.3 & 18.3 & 42.4 & 38.0 & 33.8 & 42.4 & 38.6 & 33.9 & 87.8 & 89.2 & 88.5 & 99.7 & 99.1 & 96.2 & 99.7 & 99.1 & 97.4 & 99.7 & 99.0 & 97.4 \\
    \midrule
    \multirow{4}{*}{5.}
    &  50 & 8.5 & 7.4 & 6.8 & 8.0 & 5.8 & 5.8 & 8.0 & 6.5 & 5.8 & 8.0 & 6.4 & 5.9 & 17.0 & 17.1 & 14.6 & 26.2 & 15.6 & 9.2 & 25.3 & 16.3 & 10.4 & 25.2 & 16.9 & 11.3 \\
    & 100 & 12.4 & 11.1 & 9.9 & 10.8 & 7.8 & 6.7 & 11.6 & 10.1 & 8.2 & 11.6 & 10.2 & 8.2 & 29.4 & 30.0 & 26.5 & 53.5 & 34.6 & 20.8 & 49.2 & 37.4 & 24.9 & 49.0 & 37.4 & 25.1 \\
    & 250 & 24.3 & 22.4 & 20.5 & 19.2 & 13.8 & 9.9 & 23.1 & 19.4 & 15.9 & 22.6 & 19.3 & 15.8 & 58.2 & 60.3 & 58.7 & 90.7 & 80.5 & 60.9 & 89.3 & 82.7 & 67.2 & 89.1 & 82.2 & 67.2 \\
    & 500 & 44.9 & 44.4 & 42.2 & 35.7 & 25.3 & 16.8 & 41.6 & 37.9 & 33.1 & 42.4 & 37.9 & 33.3 & 85.5 & 86.9 & 86.6 & 99.6 & 98.6 & 93.7 & 99.5 & 99.0 & 96.1 & 99.5 & 99.0 & 96.0 \\
    \bottomrule
    \end{tabular}
    }
\end{table}

    \subsection{Equality of variances: blood pressure data}\label{sec:realdata}

    This subsection revisits the equality of variances problem using data from the 2017–2018 cycle of the NHANES, a continuous survey conducted by the National Center for Health Statistics to assess the health and nutritional status of the U.S.\ civilian noninstitutionalized population \citep{NHANES2017}.
    We use the blood pressure methodology study, in which systolic blood pressure was measured under two protocols: auscultatory, performed by a physician with a mercury sphygmomanometer, and oscillometric, performed by a health technician with an oscillometric device, each producing three consecutive readings in mmHg. The six variables are therefore repeated measurements of the same physiological quantity. Restricting to participants with all six readings observed leaves $n=4{,}875$. The readings are strongly correlated, with estimated correlations around $0.96$ within the auscultatory block, $0.95$ within the oscillometric one and $0.84$ between protocols.

    Let $A_1,A_2,A_3$ denote the auscultatory readings, $O_1,O_2,O_3$ denote the oscillometric ones. We consider
    \[
    H_{AO}:\ \V\big(A_1\big)=\cdots=\V\big(O_3\big),
    \quad
    H_{A}:\ \V\big(A_1\big)=\V\big(A_2\big)=\V\big(A_3\big),
    \quad
    H_{O}:\ \V\big(O_1\big)=\V\big(O_2\big)=\V\big(O_3\big),
    \]
    with $q=6$ for $H_{AO}$ and $q=3$ for $H_A$ and $H_O$. The complete-data statistics are $T_n=26.07$ for $H_{AO}$ and $T_n=19.12$ for $H_{A}$, both rejecting at the $5\%$ level, and $T_n=5.05$ for $H_{O}$, which does not.

    We treat these participants as a fixed reference sample and impose an
    externally generated MCAR pattern, each component being retained
    independently with common probability $\varpi$.
     For each value of $\varpi$, we estimate the asymptotic noncentrality parameters in \eqref{eq:noncentr} for the AC and CC procedures using the contrast and covariance matrix estimated from the complete reference sample.  In all three testing problems, the corresponding noncentralities cross at a unique interior value $\varpi^*$, with AC being asymptotically more efficient for $\varpi<\varpi^*$ and CC for $\varpi>\varpi^*$. The resulting values are $\varpi^*=0.777$ for $H_{AO}$, $\varpi^*=0.276$ for $H_A$, and $\varpi^*=0.383$ for $H_O$. For each of several values of $\varpi$ we generate $10{,}000$ samples as in \eqref{eq:augmented_observations} keeping $n$ fixed,
    and apply AC, CC, BL and MB procedures. Since $H_O$ has little signal, with rejection frequencies below approximately $15\%$ even at $\varpi=0.9$, we do not pursue this hypothesis further. Table~\ref{tab:real_app1} reports the results for $H_{AO}$ and $H_A$. The powers reported in that table are the raw ones, that is, without applying the correction \eqref{eq:app1_lloyd}.
    
    \begin{table}[ht]
    \centering
    \caption{Average numbers of observations, and rejection percentages of the AC, CC,  BL and MB.}
    \label{tab:real_app1}

    \begin{tabular}{c cccccc cccccc}
    \toprule
    & \multicolumn{6}{c}{$H_{AO}$ \ $q=6$, \ $\varpi^{*}=0.777$} & \multicolumn{6}{c}{$H_{A}$ \ $q=3$, \ $\varpi^{*}=0.276$} \\
    \cmidrule(lr){2-7} \cmidrule(lr){8-13}
    & \multicolumn{2}{c}{$\bar N$} & \multicolumn{4}{c}{rejection \%} & \multicolumn{2}{c}{$\bar N$} & \multicolumn{4}{c}{rejection \%} \\
    \cmidrule(lr){2-3}\cmidrule(lr){4-7} \cmidrule(lr){8-9} \cmidrule(lr){10-13}
    $\varpi$ & AC & CC & AC & CC & BL & MB & AC & CC & AC & CC & BL & MB \\
    \midrule
    0.9 & 4388 & 2591 & 62.8 & 83.3 & 84.7 & 84.8 & 4388 & 3554 & 62.7 & 99.6 & 98.9 & 98.9 \\
    0.8 & 3900 & 1278 & 32.7 & 39.0 & 42.1 & 42.5 & 3900 & 2496 & 32.8 & 86.9 & 80.9 & 81.2 \\
    0.7 & 3413 &  574 & 20.6 & 17.4 & 21.9 & 22.6 & 3413 & 1672 & 20.3 & 62.1 & 54.5 & 55.4 \\
    0.6 & 2925 &  227 & 15.3 &  9.4 & 13.7 & 14.7 & 2925 & 1053 & 14.7 & 38.8 & 33.4 & 34.7 \\
    0.5 & 2438 &   76 & 10.9 &  8.1 & 10.1 & 11.5 & 2438 &  609 & 11.1 & 22.9 & 20.0 & 21.2 \\
    0.4 & 1950 &   20 &  8.0 & 15.6 & 14.5 & 17.5 & 1950 &  312 &  7.8 & 13.5 & 11.8 & 13.1 \\
    \bottomrule
    \end{tabular}
    
    \end{table}

Looking at Table \ref{tab:real_app1}, we see that the
CC procedure is more powerful at high availabilities, whereas the AC
procedure becomes more powerful as the amount of missingness increases,
which is consistent with the comparison above. The BL and MB  closely match the most powerful procedure. However, the results at low
availabilities should be interpreted with caution, since, as observed
in the previous subsection, the tests may fail to maintain their
nominal level for small effective sample sizes.

    \subsection{Coefficients of variation: simulation study}\label{sec:app2cv}\label{subsect:cv.setting}\label{sec:app:cv}
    
    Let $X_s=\big(X_s^{(1)},X_s^{(2)}\big)^\top$, $1 \leq s \leq k$, be $k$ independent populations generated as follows,
    \[
    X_s^{(\ell)}=G_{s0}+G_{s\ell},
    \qquad
    G_{s0}\sim\mathrm{Gamma}(\kappa_0,\vartheta_s),
    \quad
    G_{s\ell}\sim\mathrm{Gamma}(\kappa_1,\vartheta_s),
    \quad \ell=1,2,
    \]
    with all the summands independent, so that $X_s^{(\ell)}\sim\mathrm{Gamma}(\kappa,\vartheta_s)$, $\kappa=\kappa_0+\kappa_1$. Let $CV_s^{(l)}$ be the CV computed on the component $\ell$ of the population $s$. In this case, 
    $CV_s^{(\ell)}=\kappa^{-1/2},$ and    $\rho_x=\mathrm{Corr}\big(X_s^{(1)},X_s^{(2)}\big)=\kappa_0/\kappa$. Here, we are interested in testing 
    \[H_0: CV^{(\ell)}_i= CV^{(\ell)}_j, \quad \ell=1,2, \quad 1 \leq i,j \leq k.\]
    We take $\kappa=4$, so that $CV=0.5$, and $\vartheta_s=2s-1$: under the null the $k$ populations are genuinely different, with means $4\vartheta_s$ and standard deviations $2\vartheta_s$, but they have the same coefficients of variation. For each population, $\eta_s$ collects the two means and the two variances, estimated by $U$-statistics of degrees $1$ and $2$ with singleton index sets $A_{\ell}=\{\ell\}$, and $f(\eta_s)=\big(CV_s^{(1)},CV_s^{(2)}\big)^\top$.

    All samples are subject to componentwise missingness. Writing $\varpi_s=\big(\varpi_s^{(1)},\varpi_s^{(2)}\big)$ for the availability probabilities of population $s$, five mechanisms are considered:
    \begin{enumerate} \itemsep=0pt
    \item[Case 1.] Complete data, $\varpi_s=(1,1)$  $\forall s$.
    \item[Case 2.] Independent with $\varpi_s=(0.75,0.75)$ $\forall s$.
    \item[Case 3.] Independent with $\varpi_s=(0.9,0.6)$  $ \forall s$.
    \item[Case 4.] Dependent, generated in every population through a Gaussian copula mechanism with $\theta=0.3$ (see Section~\ref{sec:copula} of the Supporting Information).
    \item[Case 5.] Heterogeneous across populations, combining the mechanisms of the preceding cases: for $k=2$, independent $(0.75,0.75)$ and Gaussian copula ($\theta=0.3$); for $k=3$, independent $(0.9,0.6)$, independent $(0.75,0.75)$ and Gaussian copula; for $k=4$, complete data, independent $(0.9,0.6)$, independent $(0.75,0.75)$ and Gaussian copula.
    \end{enumerate}

    Table~\ref{tab:app2_level} reports the empirical sizes  at the nominal level $\alpha=0.05$ for $k\in\{2,3,4\}$, $\rho_x\in\{0.3,0.95\}$, $n_1=\cdots=n_k=n\in\{100,250,500,1000\}$. 
\begin{table}[h]
\centering
\caption{Empirical sizes (in \%) at the nominal level $\alpha=0.05$.}
\label{tab:app2_level}
\resizebox{\textwidth}{!}{
\begin{tabular}{ll|ccc|ccc|ccc|ccc|ccc|ccc|ccc|ccc}
\toprule
\multirow{3}{*}{Case} &
 &
\multicolumn{12}{c|}{$\rho_x=0.3$} &
\multicolumn{12}{c}{$\rho_x=0.95$} \\
\cmidrule(lr){3-14}\cmidrule(lr){15-26}
&
&
\multicolumn{3}{c|}{AC} &
\multicolumn{3}{c|}{CC} &
\multicolumn{3}{c|}{BL} &
\multicolumn{3}{c|}{MB} &
\multicolumn{3}{c|}{AC} &
\multicolumn{3}{c|}{CC} &
\multicolumn{3}{c|}{BL} &
\multicolumn{3}{c}{MB} \\
\cmidrule(lr){3-5}\cmidrule(lr){6-8}\cmidrule(lr){9-11}\cmidrule(lr){12-14}
\cmidrule(lr){15-17}\cmidrule(lr){18-20}\cmidrule(lr){21-23}\cmidrule(lr){24-26}
&$n\backslash k$&
$2$ & $3$ & $4$ &
$2$ & $3$ & $4$ &
$2$ & $3$ & $4$ &
$2$ & $3$ & $4$ &
$2$ & $3$ & $4$ &
$2$ & $3$ & $4$ &
$2$ & $3$ & $4$ &
$2$ & $3$ & $4$ \\
\midrule
\multirow{4}{*}{1.}
&   100 & 6.9 & 7.5 & 8.4 & 6.9 & 7.5 & 8.4 & 7.0 & 7.8 & 8.4 & 5.7 & 5.8 & 6.0 & 6.6 & 7.3 & 7.7 & 6.6 & 7.3 & 7.7 & 6.5 & 7.2 & 7.8 & 5.4 & 5.3 & 5.5 \\
&   250 & 5.4 & 6.5 & 6.7 & 5.4 & 6.5 & 6.7 & 5.6 & 6.7 & 6.8 & 5.1 & 6.0 & 5.9 & 5.8 & 5.8 & 6.5 & 5.8 & 5.8 & 6.5 & 5.9 & 5.9 & 6.5 & 5.6 & 5.2 & 5.7 \\
&   500 & 5.2 & 6.3 & 6.0 & 5.2 & 6.3 & 6.0 & 5.2 & 6.6 & 6.2 & 5.1 & 6.2 & 5.8 & 5.3 & 5.2 & 5.9 & 5.3 & 5.2 & 5.9 & 5.3 & 5.4 & 5.8 & 5.1 & 5.1 & 5.5 \\
&  1000 & 5.0 & 5.6 & 5.7 & 5.0 & 5.6 & 5.7 & 5.0 & 5.7 & 5.8 & 5.1 & 5.4 & 5.7 & 5.4 & 5.3 & 5.7 & 5.4 & 5.3 & 5.7 & 5.4 & 5.4 & 5.7 & 5.4 & 5.2 & 5.5 \\
\midrule
\multirow{4}{*}{2.}
&   100 & 6.8 & 7.6 & 9.2 & 7.4 & 8.1 & 10.2 & 7.6 & 8.4 & 10.8 & 5.2 & 5.0 & 5.9 & 5.2 & 5.7 & 6.9 & 6.7 & 7.5 & 8.4 & 6.2 & 7.1 & 8.1 & 4.3 & 4.1 & 4.2 \\
&   250 & 5.8 & 6.7 & 6.9 & 5.9 & 7.2 & 7.2 & 6.2 & 7.1 & 7.4 & 5.3 & 5.8 & 5.7 & 4.9 & 5.5 & 5.2 & 6.0 & 6.3 & 6.8 & 5.5 & 5.9 & 6.4 & 5.1 & 5.2 & 5.3 \\
&   500 & 5.5 & 6.0 & 6.1 & 5.5 & 6.0 & 6.7 & 5.6 & 6.2 & 6.9 & 5.3 & 5.7 & 6.2 & 4.8 & 5.1 & 5.0 & 5.2 & 5.8 & 6.3 & 4.8 & 5.3 & 5.8 & 4.8 & 5.3 & 5.4 \\
&  1000 & 5.2 & 5.6 & 6.1 & 5.4 & 5.7 & 5.9 & 5.3 & 5.8 & 6.3 & 5.3 & 5.6 & 5.9 & 4.8 & 4.8 & 5.2 & 5.3 & 5.5 & 5.4 & 5.2 & 4.9 & 5.4 & 5.1 & 4.9 & 5.4 \\
\midrule
\multirow{4}{*}{3.}
&   100 & 6.5 & 8.0 & 9.2 & 7.3 & 8.3 & 10.5 & 7.3 & 8.9 & 10.5 & 5.0 & 5.2 & 5.9 & 5.3 & 5.9 & 6.7 & 6.0 & 7.6 & 8.4 & 5.8 & 7.1 & 8.1 & 4.2 & 4.2 & 4.1 \\
&   250 & 6.4 & 6.7 & 7.5 & 6.3 & 7.2 & 8.6 & 6.8 & 7.3 & 8.3 & 6.0 & 6.1 & 6.4 & 4.8 & 5.6 & 5.7 & 6.0 & 6.7 & 7.2 & 5.3 & 6.3 & 6.5 & 5.0 & 5.5 & 5.6 \\
&   500 & 5.6 & 6.2 & 6.5 & 5.8 & 6.1 & 6.9 & 5.8 & 6.2 & 7.1 & 5.5 & 5.6 & 6.3 & 4.7 & 4.5 & 5.2 & 5.3 & 5.5 & 6.4 & 5.1 & 5.2 & 6.1 & 5.1 & 5.1 & 5.6 \\
&  1000 & 5.1 & 5.2 & 6.1 & 5.2 & 5.6 & 6.1 & 5.2 & 5.5 & 6.0 & 5.1 & 5.1 & 5.6 & 5.1 & 5.1 & 5.1 & 5.1 & 5.7 & 5.8 & 5.0 & 5.8 & 5.5 & 5.0 & 5.8 & 5.5 \\
\midrule
\multirow{4}{*}{4.}
&   100 & 6.5 & 7.9 & 9.3 & 7.1 & 8.6 & 10.7 & 7.1 & 9.1 & 10.9 & 4.8 & 5.3 & 6.1 & 5.1 & 6.3 & 6.6 & 6.8 & 7.6 & 8.1 & 5.9 & 7.2 & 7.7 & 4.2 & 4.3 & 4.1 \\
&   250 & 5.9 & 6.7 & 6.8 & 6.2 & 7.2 & 7.1 & 6.3 & 7.5 & 7.3 & 5.4 & 6.3 & 5.7 & 4.5 & 5.1 & 5.7 & 5.6 & 6.3 & 7.2 & 5.3 & 5.8 & 6.6 & 4.8 & 5.1 & 5.4 \\
&   500 & 5.9 & 5.7 & 6.4 & 6.2 & 6.1 & 6.9 & 6.0 & 6.1 & 7.0 & 5.5 & 5.6 & 6.3 & 5.5 & 4.9 & 5.2 & 5.8 & 5.9 & 5.9 & 5.5 & 5.4 & 5.4 & 5.2 & 5.1 & 5.2 \\
&  1000 & 5.1 & 5.1 & 5.9 & 5.0 & 5.3 & 6.0 & 5.2 & 5.4 & 6.4 & 5.1 & 5.1 & 5.9 & 4.7 & 5.2 & 5.3 & 5.0 & 5.7 & 6.0 & 5.0 & 5.4 & 5.8 & 5.0 & 5.3 & 5.7 \\
\midrule
\multirow{4}{*}{5.}
&   100 & 6.7 & 8.0 & 8.8 & 7.3 & 8.6 & 10.0 & 7.4 & 8.9 & 10.4 & 5.2 & 5.3 & 6.0 & 5.3 & 5.9 & 6.5 & 6.6 & 7.5 & 8.7 & 6.1 & 7.1 & 7.9 & 4.6 & 4.1 & 4.7 \\
&   250 & 5.6 & 6.3 & 6.8 & 6.2 & 6.9 & 7.7 & 6.0 & 6.9 & 7.8 & 5.3 & 5.7 & 6.4 & 4.9 & 5.5 & 5.3 & 5.7 & 7.2 & 6.7 & 5.3 & 6.4 & 6.4 & 4.7 & 5.6 & 5.3 \\
&   500 & 5.7 & 5.9 & 6.4 & 5.7 & 6.4 & 6.4 & 5.9 & 6.5 & 6.7 & 5.7 & 5.7 & 6.1 & 4.8 & 5.3 & 5.4 & 4.9 & 5.9 & 6.4 & 4.9 & 5.4 & 6.1 & 4.8 & 5.2 & 5.9 \\
&  1000 & 5.2 & 6.0 & 5.6 & 5.2 & 6.0 & 5.8 & 5.3 & 6.0 & 5.9 & 5.2 & 5.8 & 5.6 & 5.0 & 4.9 & 5.1 & 5.0 & 5.6 & 5.7 & 4.9 & 5.4 & 5.4 & 4.9 & 5.4 & 5.4 \\
\bottomrule
\end{tabular}
}
\end{table}

Looking at Table~\ref{tab:app2_level} we observe that for complete data and small sample sizes, the empirical size increases with $k$. The same phenomenon is observed under missingness for AC and more markedly for the CC, whose effective sample size is substantially smaller, and the BL method.  The levels of the MB procedure are in most cases closer to the target value. As $n$ increases, the empirical sizes of all methods approach the nominal level.
    
For the power study, the shape parameter of $G_{k1}$ is increased from $\kappa_1$ to $\kappa_1+0.8$, so that the first component of the $k$th population is $\mathrm{Gamma}(4.8,\vartheta_k)$ and $CV_k^{(1)}=4.8^{-1/2}=0.4564$, a $20\%$ increase of its shape, while the remaining coefficients stay at $0.5$. Table~\ref{tab:app2_powadj} reports the size-adjusted powers, obtained by applying correction \eqref{eq:app1_lloyd} as in Subsection~\ref{sec:app:vardim}.
\begin{table}[h]
\centering
\caption{Size-adjusted powers (in \%), at the nominal level $\alpha=0.05$.}
\label{tab:app2_powadj}
\resizebox{\textwidth}{!}{
\begin{tabular}{ll|ccc|ccc|ccc|ccc|ccc|ccc|ccc|ccc}
\toprule
\multirow{3}{*}{Case} &
 &
\multicolumn{12}{c|}{$\rho_x=0.3$} &
\multicolumn{12}{c}{$\rho_x=0.95$} \\
\cmidrule(lr){3-14}\cmidrule(lr){15-26}
&
&
\multicolumn{3}{c|}{AC} &
\multicolumn{3}{c|}{CC} &
\multicolumn{3}{c|}{BL} &
\multicolumn{3}{c|}{MB} &
\multicolumn{3}{c|}{AC} &
\multicolumn{3}{c|}{CC} &
\multicolumn{3}{c|}{BL} &
\multicolumn{3}{c}{MB} \\
\cmidrule(lr){3-5}\cmidrule(lr){6-8}\cmidrule(lr){9-11}\cmidrule(lr){12-14}
\cmidrule(lr){15-17}\cmidrule(lr){18-20}\cmidrule(lr){21-23}\cmidrule(lr){24-26}
&$n \backslash k$&
$2$ & $3$ & $4$ &
$2$ & $3$ & $4$ &
$2$ & $3$ & $4$ &
$2$ & $3$ & $4$ &
$2$ & $3$ & $4$ &
$2$ & $3$ & $4$ &
$2$ & $3$ & $4$ &
$2$ & $3$ & $4$ \\
\midrule
\multirow{4}{*}{1.}
&   100 & 9.7 & 9.2 & 10.0 & 9.7 & 9.2 & 10.0 & 9.8 & 9.0 & 10.0 & 10.0 & 9.6 & 10.5 & 27.6 & 25.5 & 23.8 & 27.6 & 25.5 & 23.8 & 28.0 & 25.6 & 23.5 & 28.4 & 26.3 & 24.2 \\
&   250 & 20.7 & 19.0 & 19.0 & 20.7 & 19.0 & 19.0 & 20.7 & 18.9 & 19.1 & 21.1 & 19.1 & 19.2 & 57.6 & 56.8 & 52.2 & 57.6 & 56.8 & 52.2 & 57.3 & 56.6 & 52.2 & 57.2 & 57.1 & 52.5 \\
&   500 & 36.6 & 35.0 & 36.2 & 36.6 & 35.0 & 36.2 & 36.7 & 34.3 & 35.6 & 36.4 & 34.7 & 36.0 & 86.9 & 86.4 & 82.0 & 86.9 & 86.4 & 82.0 & 86.9 & 86.1 & 82.1 & 87.0 & 86.3 & 82.3 \\
&  1000 & 65.6 & 66.7 & 65.0 & 65.6 & 66.7 & 65.0 & 65.4 & 66.5 & 65.1 & 65.2 & 67.0 & 64.6 & 99.0 & 99.0 & 98.7 & 99.0 & 99.0 & 98.7 & 99.1 & 99.0 & 98.8 & 99.0 & 99.0 & 98.8 \\
\midrule
\multirow{4}{*}{2.}
&   100 & 8.5 & 8.7 & 7.8 & 7.1 & 8.3 & 7.0 & 7.8 & 8.9 & 7.3 & 8.2 & 9.4 & 7.6 & 12.8 & 12.6 & 10.4 & 18.6 & 17.6 & 15.9 & 18.2 & 16.8 & 14.9 & 19.2 & 17.7 & 15.5 \\
&   250 & 15.7 & 14.1 & 14.7 & 12.9 & 11.9 & 12.3 & 14.6 & 14.1 & 13.9 & 14.8 & 14.5 & 14.8 & 24.5 & 23.7 & 23.6 & 38.0 & 35.5 & 33.5 & 37.1 & 35.5 & 32.6 & 37.1 & 35.1 & 32.9 \\
&   500 & 27.5 & 27.0 & 27.3 & 21.9 & 21.5 & 20.4 & 25.9 & 25.8 & 24.5 & 26.1 & 26.1 & 24.5 & 45.4 & 45.0 & 44.6 & 65.2 & 61.4 & 57.2 & 65.0 & 62.4 & 57.8 & 64.5 & 61.8 & 58.3 \\
&  1000 & 52.0 & 51.8 & 50.3 & 40.7 & 40.7 & 39.6 & 49.7 & 49.6 & 47.7 & 49.0 & 49.6 & 48.1 & 74.3 & 77.1 & 75.3 & 90.4 & 90.1 & 87.5 & 90.4 & 90.8 & 88.0 & 90.4 & 90.6 & 87.9 \\
\midrule
\multirow{4}{*}{3.}
&   100 & 9.9 & 9.2 & 9.0 & 7.5 & 8.0 & 7.4 & 8.9 & 8.8 & 8.5 & 9.1 & 9.2 & 8.7 & 13.6 & 12.9 & 12.2 & 19.7 & 16.9 & 15.7 & 19.8 & 17.6 & 15.5 & 19.6 & 17.7 & 16.2 \\
&   250 & 17.0 & 17.7 & 15.1 & 12.6 & 12.4 & 10.0 & 15.6 & 16.3 & 14.0 & 15.6 & 16.2 & 14.5 & 27.4 & 26.9 & 25.4 & 36.3 & 33.7 & 31.2 & 38.8 & 35.3 & 33.8 & 38.0 & 35.4 & 33.4 \\
&   500 & 32.8 & 31.4 & 30.6 & 20.9 & 20.5 & 19.0 & 30.4 & 29.5 & 27.8 & 30.1 & 30.0 & 27.9 & 51.3 & 52.8 & 49.9 & 62.9 & 60.4 & 56.2 & 65.9 & 64.3 & 60.5 & 65.2 & 64.0 & 60.5 \\
&  1000 & 58.5 & 62.1 & 58.9 & 39.1 & 39.2 & 37.4 & 55.9 & 58.1 & 56.4 & 55.2 & 58.8 & 56.7 & 80.3 & 82.5 & 81.7 & 89.0 & 87.9 & 85.8 & 91.5 & 90.8 & 90.1 & 91.3 & 90.6 & 89.9 \\
\midrule
\multirow{4}{*}{4.}
&   100 & 9.3 & 8.4 & 8.1 & 7.9 & 7.4 & 6.9 & 8.9 & 7.8 & 7.5 & 9.5 & 8.6 & 7.8 & 13.1 & 11.7 & 11.3 & 18.4 & 17.5 & 16.9 & 18.6 & 16.8 & 16.1 & 19.2 & 17.6 & 16.9 \\
&   250 & 15.7 & 15.0 & 14.7 & 12.2 & 11.7 & 13.2 & 14.4 & 13.2 & 14.4 & 14.7 & 13.2 & 15.1 & 26.0 & 25.6 & 22.7 & 38.5 & 36.8 & 32.0 & 37.8 & 36.8 & 32.8 & 38.0 & 36.8 & 33.1 \\
&   500 & 27.2 & 28.0 & 25.9 & 20.8 & 21.5 & 20.3 & 26.0 & 26.0 & 24.2 & 26.6 & 26.2 & 24.2 & 42.8 & 46.8 & 44.1 & 62.7 & 62.4 & 59.3 & 62.5 & 62.6 & 59.7 & 62.8 & 62.8 & 59.6 \\
&  1000 & 51.6 & 54.5 & 50.6 & 40.8 & 42.5 & 39.8 & 48.6 & 51.2 & 47.5 & 49.0 & 51.6 & 48.1 & 75.1 & 76.4 & 76.3 & 90.9 & 89.4 & 87.4 & 90.4 & 89.5 & 87.8 & 90.2 & 89.3 & 87.7 \\
\midrule
\multirow{4}{*}{5.}
&   100 & 8.7 & 9.5 & 8.8 & 7.7 & 8.1 & 7.9 & 8.4 & 8.8 & 8.4 & 8.3 & 8.9 & 8.7 & 13.0 & 12.7 & 13.1 & 19.5 & 17.6 & 16.5 & 18.9 & 17.3 & 16.6 & 18.4 & 18.2 & 16.7 \\
&   250 & 16.2 & 16.2 & 15.9 & 12.5 & 11.9 & 12.9 & 14.8 & 14.5 & 14.4 & 14.8 & 14.8 & 14.5 & 25.2 & 24.9 & 28.2 & 38.9 & 33.3 & 34.8 & 38.4 & 34.0 & 35.5 & 38.7 & 33.9 & 35.8 \\
&   500 & 27.1 & 29.1 & 27.1 & 21.0 & 20.4 & 22.0 & 25.0 & 26.0 & 25.7 & 24.2 & 27.0 & 25.9 & 44.9 & 47.3 & 50.1 & 66.1 & 62.5 & 59.3 & 64.7 & 63.6 & 60.9 & 64.0 & 63.5 & 60.6 \\
&  1000 & 51.3 & 53.4 & 54.6 & 40.5 & 40.1 & 43.4 & 48.4 & 50.7 & 52.0 & 48.5 & 50.4 & 52.1 & 74.5 & 78.5 & 81.2 & 90.7 & 89.3 & 88.6 & 90.4 & 90.1 & 90.0 & 90.3 & 89.9 & 89.8 \\
\bottomrule
\end{tabular}
}
\end{table}
Looking at this table, we observe that as before, in all cases the BL and MB procedures lie between AC and CC, and they are generally closer to the method with better performance.

    \subsection{Coefficients of variation across sex}\label{sec:rd:app2}
    
     Here we consider the dataset in Subsection \ref{sec:realdata}. Specifically, we focus on one of the eleven urinary metals in the panel, measured in $\mu$g/L, and total serum cholesterol, measured in mg/dL. The analysis is restricted to the $4{,}937$ participants aged $20$ or over for whom cholesterol was recorded, comprising $2{,}377$ men and $2{,}560$ women. Let $X_i=\big(X_i^{(1)},X_i^{(2)}\big)^\top$, where $X_i^{(1)}$ and $X_i^{(2)}$ denote the urinary concentration of the metal and total cholesterol, respectively, for a participant of sex $i$, with $i=1$ for men and $i=2$ for women. We test the equality of the coefficients of variation of the two components between men and women,
\[
H_0:\quad CV_1^{(1)}=CV_2^{(1)}
\quad\text{and}\quad
CV_1^{(2)}=CV_2^{(2)}.
\]
Cholesterol is fully observed, whereas the urinary metal was measured on a randomly selected third of the participants, yielding $\widehat{\pi}^{(1)}=0.328$ overall ($0.334$ among men and $0.322$ among women). Thus, the missingness mechanism is MCAR by construction. Consequently, the AC estimator uses all $n_i$ participants for cholesterol and the $N_i^{(1)}$ participants with the metal measured, whereas the CC estimator uses only the $N_i^{(1)}$ complete cases for both components. In particular, the CC analysis discards approximately two thirds of the available cholesterol measurements. The analysis is repeated for each of the eleven metals. We applied Holm-Bonferroni method to adjust the obtained $p$-vales, which  are reported in Table \ref{tab:rd:app2}.
\begin{table}[t]
\centering
\caption{Holm-Bonferroni adjusted $p$-values (in \%) for testing equality of coefficients of variation across sex for total cholesterol and eleven urinary metals using the AC, CC, BL and MB procedures.}
\label{tab:rd:app2}
\resizebox{\textwidth}{!}{
\begin{tabular}{lccccccccccc}
\toprule
Metal & Antimony & Barium & Cadmium & Cesium & Cobalt & Lead & Manganese & Molybdenum & Thallium & Tin & Tungsten \\
\midrule
$p_{AC}$ & $4.80$ & $7.06$ & $7.06$ & $0.44$ & $7.06$ & $4.85$ & $4.85$ & $7.06$ & $1.19$ & $7.06$ & $6.63$ \\
$p_{CC}$ & $45.67$ & $100$ & $100$ & $0.09$ & $100$ & $45.67$ & $45.67$ & $100$ & $22.44$ & $100$ & $100$ \\
$p_{BL}$ & $7.29$ & $11.44$ & $11.22$ & $0.44$ & $11.44$ & $7.44$ & $7.49$ & $11.44$ & $2.20$ & $11.44$ & $11.22$ \\
$p_{MB}$ & $41.20$ & $41.20$ & $37.08$ & $0.22$ & $41.20$ & $41.20$ & $41.20$ & $41.20$ & $2.20$ & $41.20$ & $41.20$ \\
\bottomrule
\end{tabular}
}
\end{table}
Since total cholesterol and urinary metal concentrations are independent, Remark~\ref{rem:kcomparison} and Corollary~\ref{cor:indep} imply that the AC test has local asymptotic power at least as large as that of the CC test. This is consistent with the results in Table~\ref{tab:rd:app2}.

\section{Conclusions and further research}
\label{sec:conclusions}

Whether to base inference on the AC or CC estimators is a long-standing practical dilemma. While several papers have addressed specific problems in point estimation, hypothesis testing has received comparatively little attention in the statistical literature. This paper develops the tools needed to address this problem in a fully nonparametric framework, for parameters that can estimated through  functions of vectors of $U$-statistics computed from randomly overlapping subsets of the sample.

We derive the joint asymptotic normality of the AC estimators, allowing for component-specific normalizations and availability probabilities that may vary with the sample size. The limiting covariance matrix factorizes into the complete-data covariance matrix and a matrix determined solely by the availability pattern. The CC approach can be viewed as a special case of this framework. To conduct further inference, a consistent estimator of the variance of the asymptotic distribution is required. Since the jackknife does not recover the factor associated with the missingness mechanism, we estimate this factor separately, yielding a consistent variance estimator.

These results provide the theoretical tools needed to compare AC and CC based Wald-type tests. This comparison, however, is specific to each testing problem and depends, among other factors,  on the  limiting covariance matrix, which limits its practical usefulness. Rather than requiring a choice between the two approaches, we study a test based on the maximum of the AC and CC test statistics,  whose distribution can be consistently approximated using a bootstrap in the limit and a multiplier bootstrap estimators. 
In simulations, the resulting test typically exhibited power between that of the AC and CC tests, generally closer to that of the more powerful approach.


Several extensions are possible. (i) The Wald statistics require a generalized inverse of the estimated covariance matrix, which can be computationally demanding, and, as the simulations show, the resulting tests tend to be liberal in small samples. In this context, ANOVA-type statistics (see e.g.,  \cite{Brunner1997,articulo2}) which avoid the inverse at the price of a non-pivotal limit (a point with these tests statistics is that their null distribution  must be numerically approximated), emerge as a natural alternative, and it would be interesting to explore whether the tests based on them improve the finite-sample performance.  (ii) The dimension of the parameter has been kept fixed. Extending the joint limit theory to a growing number of components, along the lines of \cite{YIN2021104797} and \cite{BHATTACHARJEE2026105672}, would cover the settings in which the advantage of the available-case approach is largest, since the proportion of complete cases decreases fastest there. (iii) In the $k$-sample problem the number of populations has also been kept fixed. The comparison of $k$ $U$-statistics with $k$ growing with the sample sizes has been studied for complete data in \cite{articulo1}, and extending that analysis to available-case estimators deserves a separate study.

\section*{Supporting Information}
The Supporting Information (PDF) contains proofs of the theoretical results,
classical results on $U$-statistics, and details of the Gaussian copula
missingness mechanism.

\section*{Acknowledgements}

This research has been financed by research project PID2022-137818OB-I00 (Ministerio de
Ciencia, Innovaci\'on y Universidades, Spain). The authors thank IMUS-Maria de Maeztu
grant CEX2024-001517-M - Apoyo a Unidades de Excelencia Mar\'ia de Maeztu for supporting
this research, funded by MICIU/AEI/ 10.13039/501100011033. 

The authors also acknowledge the use of ChatGPT (OpenAI) as a tool for
assistance with mathematical derivations, checking arguments, and
improving the exposition during the revision of the manuscript. All
mathematical statements and conclusions were independently assessed
and remain the responsibility of the authors.

\bibliographystyle{apalike}
\bibliography{referencias.bib}

\end{document}